\documentclass[epsfig]{article}
\usepackage{graphicx}
\usepackage{amsmath}

\input epsf.sty
\input{tcilatex}
\makeatletter \@addtoreset{equation}{section}
\renewcommand{\theequation}{\thesection.\arabic{equation}}

\begin{document}

\title{\rightline{\mbox {\normalsize {Lab/UFR-HEP/0209}}} \textbf{NC Effective
Gauge Model for \\ Multilayer FQH States}}
\author{Aziz El Rhalami and El Hassan Saidi\thanks{%
E-mail: H-saidi@fsr.ac.ma} \\
\textit{Lab/UFR-High Energy Physics, Physics Department,}\\
\textit{\ Faculty of Science, P.O Box 1014, Avenue Ibn Battouta, Rabat,
Morocco.}}
\maketitle

\begin{abstract}
We develop an effective field model for describing FQH states with
rational filling factors that are not of Laughlin type. These
kinds of systems, which concern single layer hierarchical states
and multilayer ones, were observed experimentally; but have not
yet a satisfactory non commutative effective field description
like in the case of Susskind model. Using $D$ brane analysis and
fiber bundle techniques, we first classify such states in terms of
representations characterized, amongst others, by the filling
factor of the layers; but also by proper subgroups of the
underlying $U\left( n\right) $ gauge symmetry. Multilayer states
in the lowest Landau level are interpreted in terms of systems of
$D2$ branes; but hierarchical ones are realized as Fiber bundles
on $D2$ which we construct explicitly. In this picture, Jain and
Haldane series are recovered as special cases and have a
remarkable interpretation in terms of Fiber bundles with specific
intersection matrices. We also derive the general NC commutative
effective field and matrix models for FQH states, extending
Susskind theory, and give the general expression of the rational
filling factors as well as their non abelian gauge symmetries.
\end{abstract}
\bigskip
\begin{quote}
{\bf Keywords} \qquad: {\it Multilayer and FQH hierarchies; Branes
and fiber bundles on branes, NC non abelian Chern Simons gauge
theory, Matrix model}
\end{quote} \thispagestyle{empty}
\newpage
\thispagestyle{empty}
\tableofcontents
\newpage
\setcounter{page}{1}

\section{Introduction}

\qquad Susskind proposal \cite{1} that Non-Commutative (NC) Chern-Simons
gauge theory on the $(2+1)$ space provides a natural framework to study the
Laughlin state of filling factor $\nu _{L}=\frac{1}{k}$, $k$ a positive odd
integer. This proposal \ has opened a new way to deal with the effective
field models of Fractional Quantum Hall (FQH) fluids and offered possible
interpretations in terms of $D$ brane solitons of type $II$ superstring
theory \cite{2,3,4}. Since this important development, an intensive interest
has been given to explore further this remarkable issue and several basic
results has been derived. A regularized version of the Susskind NC effective
field theory using finite dimensional matrix model techniques has been
introduced in \cite{5} to study FQH droplets. There, it has been shown that
consistency requires \ the introduction of an extra field, the polychronakos
field, which is a regulator field playing an important role at the quantum
level \cite{6,8}. Along with these developments, it has been also
conjectured that a specific assembly of a system of $D0$, $D2$ and $D6$
branes and $F1$ strings, stretching between $D2$ and $D6$, has a low energy
dynamics similar to the fundamental state of FQH systems \cite{9,10}. For
other applications and issues see \cite{11,12}.

\qquad In Susskind NC model, the non commutativity parameter $\theta $ of
the co-moving plane coordinates is related to the filling factor $\nu _{L}$
and then to the Chern-Simon effective field coupling $\lambda _{CS}$ as $%
\theta \mathcal{B}_{ex}\nu _{L}=\nu _{L}\lambda _{CS}=1$; where $\mathcal{B}%
_{ex}$ is the external magnetic field. This relation, which get quantum
corrections, shifting the level $k$ of the CS gauge theory \cite{6,7,13},
has been used in \cite{14,15,16} to approach a specific class of states that
are not of Laughlin kind; that is FQH states with rational values type $\nu
_{2}^{\left( h\right) }=\frac{k_{1}+k_{2}}{k_{1}k_{2}}$ with $k_{1}$and $%
k_{2}$\ positive odd integers. These kinds of states, which are just the
leading elements of general ones having the filling factor taking general
rational values, belong to two kinds of FQH\ systems: (1) multilayer FQH
systems in the ground state and beyond and (2) generic levels of the so
called hierarchical series.

\qquad Like for the Laughlin ground state, FQH states with general rational
values of the filling factor, such as, the well known $\frac{2}{5}$, $\frac{2%
}{3}$,$\frac{3}{7}$,...; were observed experimentally several years ago; but
are not covered by the Susskind NC theory which, due to quantization and
unitarity conditions, requires that $1/\nu $ should be positive integer.

In \cite{14}, see also \cite{15,16}, an extension of Susskind NC field model
to cover hierarchical states has been approached by noting that some FQH
hierarchical states, at given levels $n$, can be usually decomposed as a
particular sum over $n$ Laughlin states built in a recurrent manner. In
Haldane hierarchy, this splitting feature was first noted on the leading
elements of the series such as $\frac{2}{5}$ and $\frac{3}{7}$, which have
the remarkable decomposition
\begin{eqnarray}
\frac{2}{5} &=&\frac{1}{3}+\frac{1}{15},  \notag \\
\frac{3}{7} &=&\frac{1}{3}+\frac{1}{15}+\frac{1}{35},
\end{eqnarray}
allowing to interpret them as bounds of Laughlin states. Here we will show
that this splitting is in fact valid at any level $n$ and follows due to a
remarkable exact mathematical result of the continuous fraction $\left[
p_{1}...p_{n}\right] $; which ensures that the level $n$ of Haldane series $%
\nu _{n}^{\left( Hal\right) }$ can be usually brought to the form
\begin{equation}
\mathrm{\nu }_{n}^{\left( Hal\right) }=\frac{1}{k_{1}}+\frac{1}{k_{2}}+...+%
\frac{1}{k_{n}},
\end{equation}
where $k_{i}$ are some specific odd integers to be computed explicitly in
section 5. Similarly, Jain hierarchical series may, roughly speaking, be
also thought of as given by the special decomposition,
\begin{equation}
\mathrm{\nu }_{n}^{\left( Jain\right) }=\frac{1}{2np\pm 1}+\frac{1}{2np\pm 1}%
+...+\frac{1}{2np\pm 1}.
\end{equation}
Analogous analysis may be written down as well for this sequence; for
details see section 5 eqs(5.13-17). What interest us from this brief
presentation is note that in both decompositions (1.1-.2) and (1.3), one
sees that the level $n$ of hierarchy is also the number of filled lowest
Landau levels, (\textsl{LL}$_{1},...,$\textsl{LL}$_{n}$ for short) and
moreover each \textsl{LL}$_{i}$ behaves as a kind of Laughlin state with
filling factor $1/k_{i}$; $k_{i}$ odd integer.

\qquad Despite these partial results and others established in recent
literature \cite{15,16,17,18,19,20}, there are however basic questions,
regarding states with rational filling factors, that remain without
convincing answers. Besides the non commutative non abelian $U\left(
n\right) $ Chern Simons gauge model describing multilayer states with
filling factors
\begin{equation}
\mathrm{\nu }_{U\left( n\right) }=n/k,
\end{equation}
where $k$ is the level of the Chern Simons model, no consistent NC field
theoretical construction extending Susskind NC model has been obtained
neither for hierarchical states, nor for multilayer systems. To this lack,
one should also add an other notable one concerning the classification of
FQH states with rational value\footnote{%
By rational values of the filling factor $\mathrm{\nu }$, we mean all $%
\mathrm{\nu }$'s that have the form $\mathrm{\nu =n/q,}$ with \textrm{n} and
\textrm{q} prime integers. The Laughlin fraction $\mathrm{\nu }_{L}\mathrm{%
=1/k}$ is a special case which has been extensively studied and is quite
well understood.}.\ For example, if one considers state with $\mathrm{\nu }%
=2/5$ and try to list all possible FQH systems in which it can appear, one
sees that there is a variety of possibilities: To write down this list, note
first of all that such states may appear into two kinds of system: (1)
multilayer system and (2) single layer hierarchical states, offering by the
occasion the first ingredient in the classification.

\textsl{Multilayer states}

If one focuses on the state $\mathrm{\nu }=2/5$ and considers that the two
layers $\mathcal{L}_{1}$ and $\mathcal{L}_{2}$ of the system are taken in
the ground state; then we have the following natural representations:

(a) The state $\mathrm{\nu }=2/5$ realized as an irreducible representation;
like in NC non abelian $U\left( 2\right) $ Chern Simons gauge model with a
level $k=5$. Here, the two layers $\mathcal{L}_{1}$ and $\mathcal{L}_{2}$
should be completely symmetric.

(b) As a reducible state $1/5+1/5$ like in uncoupled NC abelian $U\left(
1\right) \times U\left( 1\right) $ Chern Simons model with equal levels $%
k_{1}=k_{2}=5$. This model is expected to be obtained from the previous
representation after breaking of the $U\left( 2\right) $ gauge symmetry and
integrating out massive modes.

(c) As a reducible state $1/3+1/15$ like in uncoupled NC abelian $U\left(
1\right) \times U\left( 1\right) $ Chern Simons model with different
levels;\ i.e $k_{1}=3$ and $k_{2}=15$. Here the quantum properties of the
two layers $\mathcal{L}_{1}$ and $\mathcal{L}_{2}$ are different and this
representation seems to have nothing to do with the original NC non abelian $%
U\left( 2\right) $ Chern Simons model. One expects then to get multilayer
effective field realizations other than the NC non abelian Chern Simon gauge
model. We will show later on, that at level $n=2$ for instance, a possible
generalization $\mathrm{\nu }_{U\left( 1\right) ^{2}}$ of eq(1.4) is,
\begin{equation}
\mathrm{\nu }_{U\left( 1\right) ^{2}}=\frac{k_{1}+k_{2}}{k_{1}k_{2}-l^{2}}
\end{equation}
For $k_{1}\neq k_{2}$, this equation corresponds to the $U\left( 1\right)
^{2}$ model describing two parallel layers with different individual filling
factors; $1/k_{1}$and $1/k_{2}$, with $k_{1}$ and $k_{2}$ odd integers.\
Interactions between layers shift the inverse of the free filling factor by\
the amount $l^{2}/k_{1}k_{2}$. Moreover, this relation allows to recover the
Jain sequence, which is obtained from (1.5), by looking for integer
solutions of the following eq
\begin{equation}
l=\frac{1}{2}\sqrt{\left[ 2k_{1}-\left( 4p+1\right) \right] \left[
2k_{2}-\left( 4p+1\right) \right] -\left( 4p+1\right) ^{2}};\quad l\in Z^{+}
\end{equation}
where the integer $p$ is as in eq(1.3). The formula (1.5) we have given
above is in fact the second simplest example of a more general result to be
established in subsection 3.1, eq(3.46); expressing the filling fraction $%
\mathrm{\nu =Tr}\left[ \mathrm{\kappa }^{-1}\right] $ of multilayer FQH
states in terms of the inverse of a hermitian matrix $\mathrm{\kappa }$\ of $%
GL\left( n;\mathbf{Z}^{+}\right) $ with odd integer diagonal terms. It
permits to recover all possible picture including the non abelian model
which correspond to the solutions of the constraint eq $U\mathrm{\kappa
=\kappa }U$, where $U$\ is are $n\times n$\ matrix describing the gauge
transformations.

\textsl{Single Layer hierarchical states}

In the case of a single layer $\mathcal{L}$ and considering usually the
state $\mathrm{\nu }=2/5$, one may also write down a list of possible
realizations;

(d) The state $\mathrm{\nu }=2/5$ realized as an irreducible representation $%
2/5$, like in the Lopez and Fradkin model \cite{21},

(e) As a reducible state $1/5+1/5$ like in Jain series eq(1.3),

(f) As a reducible state $1/3+1/15$ like Haldane series decomposition
(1.1-2).

In addition to these list of representations, one should also add those
hierarchical states with rational values of filling factors living on
multilayer systems; that is systems with several layers taken outside the
ground state.

\qquad In this paper, we present a unified effective field model for
studying FQH states with rational filling fraction that come either from
multilayer systems, hierarchical states of a single layer or again
hierarchical states in multilayer systems. Our way of doing is mainly
motivated by similarities between FQH liquids and $D$ brane systems of type $%
II$ string theory \cite{pol}. Using these tools and others geometric ones,
we develop a general effective field model for FQH states with rational
filling factors. More precisely, we use ideas borrowed from $D$ brane
physics \cite{hanany} and fiber bundles, \textsl{F}$\left( \cup D2\right) $
on a set of $D2$ branes, to study FQH liquids with rational filling factors.
As a result, we obtain the general NC effective field and matrix theories
modeling FQH states with generic rational filling factors. The generating
functional action of our effective field model, which has different sectors,
reads formally, as
\begin{equation}
\mathcal{Z}\left[ A\right] =\int \prod_{\text{{\small massive fields}}}%
\mathcal{D}\left[ A\right] \exp -i\left( \int d^{3}y\sum_{\mathcal{L}%
_{a}}\sum_{\QTR{sl}{LL}_{i_{a}}}\mathrm{L}\left( A\right) \right) ,
\end{equation}
where $\left\{ \cup _{a=1}^{n}\mathcal{L}_{a}\right\} $ stands for a system
of $n$ layers $\mathcal{L}_{a}$\ interpreted as a set of \ n parallel $D2$
branes, and where $\left\{ \cup _{i_{a}=1}^{m_{a}}\QTR{sl}{LL}%
_{i_{a}}\right\} $ is the set of the $m_{a}$ filled lowest Landau level
associated with the $a-th$ layer $\mathcal{L}_{a}$. In this relation, $%
\mathrm{L}\left( A\right) $ is a lagrangian density which reads, for $%
U\left( n\right) $\ gauge invariant model, as in eq(7.6-7)\ and, in the case
$n=1$ and $m_{1}=1$, reduces to the well known Susskind NC field model for
Laughlin state. This functional action exhibits, amongst others, the
following features.

\begin{itemize}
\item  Extends, to multilayer states and hierarchical ones, the original
Susskind NC field theory and the Susskind-Polychronakos regularized matrix
model initially obtained for the case of a single layer in the ground state.
Our model is general; it contains as well the single layer states belonging
to Jain and Haldane series as particular ones.

\item  Answers remarks made in \cite{21}, regarding consistency of Wen-Zee (
WZ) model for hierarchy \cite{22,23,23'}; in particular the point concerning
the stability of the WZ conserved currents. Though, we support the arguments
given in \cite{21}, we give nevertheless validity conditions under which WZ
effective field approach works. We also work out explicitly the NC field and
matrix models generalizing such ideas.

\item  Presents a framework where single layer hierarchical states with
rational filling factors and the ground configuration of multilayers as well
as multilayer hierarchical states are treated in a unified way, in perfect
agreement with Susskind basic idea \cite{1} and non abelian symmetries of
parallel branes.
\end{itemize}

\qquad The presentation of this paper is as follows: In section 2, we
describe, in two subsections, FQH states with rational filling factors,
belonging to multilayer systems on one hand and on the other hand to a
single layer hierarchical states. The multilayer system is viewed as
represented by $n$ parallel $D2$ branes located at the positions $%
y^{3}=d_{a} $, in the direction of the external magnetic field $\mathcal{B}%
_{ex}$. The single layer hierarchical states are interpreted as fiber
bundles on $D2$ branes. Here also, we fix some terminology and convention
notations. In section 3, we study commutative effective field models for
multilayer states in the lowest Landau levels. We distinguish between
several gauge fields models with abelian and non abelian gauge symmetries.
In this regards, we show that only layers with same filling factor; say $%
k_{1}=k_{2}=...=k_{n}=k$; which can lead to a $U\left( n\right) $ non
abelian symmetry; otherwise the $U\left( n\right) $ symmetry is broken down
to subgroups depending on the number of equal $k_{a}$'s one has. In case
where all $k_{i}$'s are different, the effective $U\left( 1\right) ^{n}$
abelian gauge field models we obtain are of two types: (i) either without
interactions; i.e $k_{ab}=0$ for $a\neq b$, and then the total filling
factor is given by the sum of the individual filling factors or (ii) having
interactions; i.e $k_{ab}\neq 0$ for $a\neq b$, as in Wen-Zee theory for
single layer hierarchical states. In this case, the filling factor have a
general form given by eq(3.46), which contain Haldane and Jain series as
special cases. We also use this occasion to give a classification of the
various type of the generalizations of Laughlin wave functions one
encounters in FQH literature and take the opportunity to complete some
partial results in this matter. In section 4, we study the NC non abelian
effective gauge field and matrix models for multilayers states where the
usual NC non abelian $U\left( n\right) $ Chern Simons gauge theory appears
as just the most symmetric representation. The other less symmetric
representations are also studied. In section 5, we study the Wen-Zee
effective field theory for single layer states with rational filling factors
and review some aspects regarding Jain and Haldane hierarchies, which are
now viewed as two special fiber bundles whose explicit realizations are
given in subsections 5.2 and 5.3. In section 6, we give the NC gauge fields
and matrix models for single layer hierarchical states and in section 7, we
give a discussion concerning hierarchy in multilayer systems and make a
conclusion..

\section{ FQH Hierarchies and Mutilayer\ Systems}

\qquad Experiments on Hall systems showed the existence of stable states at
critical values of filling factor $\nu $ taking in general rational values;
the familiar $\mathrm{\nu }=\frac{1}{3},$ $\frac{2}{5}$, $\frac{3}{7}$ are
examples amongst many others. Laughlin constructed an explicit trial wave
function to explain QH state partially filled with $\mathrm{\nu }=\frac{1}{m}
$ where $m$ odd integer. He argued that the elementary excitations from the
stable states are quasiparticles with fractional electric charge $q=\pm \nu
e $ and obey a generalized statistics. In this scheme, electrons are thought
of as a kind of condensate of $m$ quasiparticles and occupy the lowest
Landau level ( LLL ). Recently Susskind completed this model by showing that
the right $\left( 1+2\right) $\ dimension effective field theory that
describe Laughlin state is a NC Chern Simons $U\left( 1\right) $ gauge model
on Moyal surfaces\footnote{%
The surface we will be considering in this paper is, roughly speakind, the
real two plane. However most of the results we will obtain may be also
extended naturally to other two dimension real geometries with and without
boundaries, such as the strip, disc, cylinder, two sphere and the torus. In
addition to boundary effects, one should also be aware about gauge
symmetries involving a parallelism condition on layers.}.

\qquad To describe the states with general values of the filing factor; say
of the form $\mathrm{\nu }=\frac{n}{p}$, there is no standard method to
follow; but rather different approaches one can use to study such states.
These ways have a common denominator; as all of them are based on the
Laughlin model. Motivated by the recent developments \cite{14,15,16} based
on Susskind model; its regularized finite dimensional matrix formulation as
well as results on hierarchical states using exact algebraic feature on the
continuous fraction, we explore in this section the main representations of
FQH systems that lead to rational values of the filling fraction. We also
fix our terminology regarding the correspondence between layers and $D2$
branes, on one hand, and Landau levels and fiber bundles on the other hand.

\subsection{Multilayer Representation}

\qquad The multilayer FQH states with rational filling factors, we will be
considering here are obtained from a system of $\mathcal{N}_{e}=\left(
N_{1}+...+N_{n}\right) $ electrons moving in a set of $n$ parallel layers $%
\mathcal{L}_{a}$ in presence of an external \ constant and orthogonal
magnetic field $\mathcal{B}_{ex}$ with a constant flux $\int \mathcal{B}%
_{ex}=\mathcal{N}_{\phi }$. As a convention of notation, we will denote this
system as,
\begin{equation}
\left\{ n\mathcal{L}\equiv \cup _{a=1}^{n}\mathcal{L}_{a};\quad \mathcal{L}%
_{a}\Vert \mathcal{L}_{b}\right\} .
\end{equation}
Note that parallelism of layers should be understood as a local condition.
If we let $\left\{ \mathcal{O}_{a}^{I};\quad I=1,2,...\right\} $ be
topological basis of local open sets covering the $\mathcal{L}_{a}$ layers
\begin{equation*}
\mathcal{L}_{a}=\cup _{I\geq 1}\mathcal{O}_{a}^{I};\quad \mathcal{O}%
_{a}^{I}\cap \mathcal{O}_{a}^{I}\neq \emptyset ,
\end{equation*}
then $\mathcal{L}_{a}$ is said parallel to $\mathcal{L}_{b}$ should be
thought of as an open set $\mathcal{O}_{a}^{I}$ of $\mathcal{L}_{a}$ is
parallel to an open set $\mathcal{O}_{b}^{J}$\ of $\mathcal{L}_{b}$. In this
paper, we will simplify the presentation by focusing our analysis directly
on layers. Moreover since the $N_{a}$ electrons on the layer $\mathcal{L}%
_{a} $ may be sitting in different Landau levels $\left\{ \QTR{sl}{LL}%
_{i}\right\} ,$ we will add, when necessary, an extra index to implement
this feature into the formalism. Thus $\mathcal{L}_{a}^{i_{a}}$ denote a FQH
configuration of electrons in $i_{a}-th$ Landau levels of the layer $%
\mathcal{L}_{a}$. Later on, we will give a representation of this
configuration in terms of a fiber bundle on $\mathcal{L}_{a}$; that is,
\begin{equation}
\mathcal{L}_{a}^{i_{a}}\equiv \left( \QTR{sl}{LL}_{i_{a}}\QTR{sl}{;}\mathcal{%
L}_{a}\right) \equiv \QTR{sl}{F}_{i_{a}}\left( \mathcal{L}_{a}\right)
\end{equation}
For the moment let us fix our attention on the lowest Landau level and
suppose that all layers $\mathcal{L}_{a}$ of the system are of Laughlin
type; i.e with filling factors $\nu _{a}=1/k_{a}$. Since the FQH ( patches
of ) layers are mainly two dimensional real surfaces, it is natural to
assimilate the $\left\{ n\mathcal{L}\right\} $ system to an assembly of $n$
parallel $D2$ branes parameterized by the local ( patches ) coordinates
\begin{eqnarray}
nD2 &\sim &\left\{ y_{a}^{\mu }=\left( y^{1},y^{2},y^{3}=d_{a}\right) ;\quad
a=1,...,n\right\} ;  \notag \\
\mathcal{B}_{ex} &=&\varepsilon ^{3\alpha \beta }\text{ }\partial _{\alpha }%
\mathcal{A}_{\beta }\equiv \varepsilon ^{30\alpha \beta }\text{ }\partial
_{\alpha }\mathcal{A}_{\beta },
\end{eqnarray}
where $\left| d_{a}-d_{b}\right| $\ is the distance separating the pair of
layers $\left( \mathcal{L}_{a},\mathcal{L}_{b}\right) $, $\varepsilon
^{3\alpha \beta }$ the completely three dimension antisymmetric tensor and
where $\mathcal{A}_{\beta }$ is the external vector potential; see figure 1.
Since electrons moving in this system have intra an inter layers
interactions, one may use the $\left| d_{a}-d_{b}\right| $ layers
inter-distances and D branes symmetries to classify the various possible
models.
\begin{figure}[tbh]
\begin{equation*}
\epsfxsize=7cm \epsffile{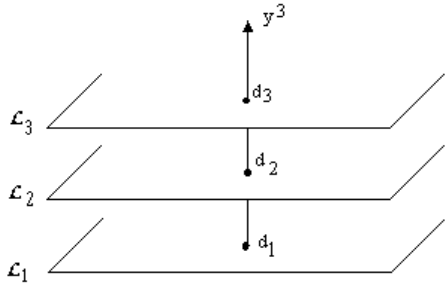}
\end{equation*}
\caption{\textit{This figure represents a system of }$n$ $\left( n=3\right) $%
\textit{\ parallel layers }$\left\{ \mathcal{L}_{a}\right\} $ \textit{normal
to the external magnetic field} $\mathcal{B}_{ex}=\protect\varepsilon ^{3%
\protect\mu \protect\nu }\partial _{\protect\mu }\mathcal{A}_{\protect\nu }$%
. \textit{\ These layers are represented by }$n$\textit{\ parallel }$D2$%
\textit{\ branes filling the plane }$(y^{1},y^{2})$\textit{\ located at the
positions }$y^{3}=d_{c};$ $c=1,...,n;$\textit{\textit{\ } with spacings }$%
d_{ab}=d_{a}-d_{b}$\textit{. These layers may be either distant enough from
each others as in} {\ $U(1)^{n}$ }\textit{\ gauge theory for instance or
very close, eventually coincident. They may also have different filling
factors.}}
\label{figure 1}
\end{figure}
According to the values of $\left| d_{a}-d_{b}\right| $, one can show that
the number of configurations of $\left\{ n\mathcal{L}\right\} $ is linked to
the subgroups of the $U\left( n\right) $ symmetry. Two special
configurations of the layers are those associated with the two following
extreme cases: (i) the case where all $D2$ branes are distant enough from
each others; i.e,
\begin{equation}
\left| d_{a}-d_{b}\right| >>>1,\quad \forall a,b=1,...,n.
\end{equation}
In this situation, layers interactions carried by massive modes may be
ignored and, roughly speaking, the underlying symmetry of the effective
Chern Simons model describing the FQH multilayer system is $U\left( 1\right)
^{n}$. (ii) The other extreme case deals with the situation where all $D2$
branes coincide, ie
\begin{equation}
\left| d_{a}-d_{b}\right| <<<1,\quad \forall a,b=1,...,n.
\end{equation}
This FQH configuration is described by an effective Chern Simons model with
a non abelian $U\left( n\right) $ gauge invariance. The remaining situations
correspond to the various systems with gauge invariances given by subgroups
of $U\left( n\right) $.

\subsection{Hierarchical Representation}

\qquad We start by noting that there are two kinds of hierarchical FQH
states; those involving several lowest Landau levels $\left\{ \cup _{i=1}^{m}%
\QTR{sl}{LL}_{i}\right\} $ of a given single layer $\mathcal{L}$ in the
sense of the decompositions eqs(1.1-3); and those implying various Landau
levels $\left\{ \cup _{a=1}^{n}\cup _{i_{a}=1}^{m_{a}}\QTR{sl}{LL}%
_{i_{a}}\right\} $ of a multilayer system $\left\{ \cup _{a=1}^{n}\mathcal{L}%
_{a}\right\} $. \ The second are naturally more general. Here we shall
describe the first kind of states; but later on we shall give the general
result.

\subsubsection{Case of a single Layer}

\qquad As we have said, FQH states with filling factor that are not of
Laughlin type exist also for the case of one layer $\mathcal{L}$. This is
the case for instance of the so called hierarchical states belonging to Jain
and Haldane series. In the picture; one layer $\mathcal{L}$ $%
\longleftrightarrow $ one $D2$ brane,
\begin{equation}
\mathcal{L}\longleftrightarrow \mathcal{D}2,  \label{cor1}
\end{equation}
where one has only one gauge field describing the displacement of the fluid,
it seems a priori not obvious to describe such kind of states using Chern
Simons $U\left( 1\right) $ gauge field model as also noted in \cite{21}. In
this section we want to present two scenarios to overcome this difficulty.
The first one is based on: (1) abandoning the correspondence; one layer $%
\mathcal{L}$ $\longleftrightarrow $ one $D2$ brane in profit of one Landau
level $\QTR{sl}{LL}$ $\longleftrightarrow $ one $D2$ brane,
\begin{equation}
\QTR{sl}{LL}\longleftrightarrow D2.
\end{equation}
In this way of viewing things, the problem is solved since we have several
gauge fields at hand and so the brane multilayer representation $\left\{
\cup _{a=1}^{n}\mathcal{L}_{a}\right\} $ considered in subsection 2.1,
translates completely to a multi Landau levels representation $\left\{ \cup
_{a=1}^{n}\QTR{sl}{LL}_{a}\right\} $. \ The results one gets are quite
similar to those one obtains for the multilayer system; it suffices to make
the substitution
\begin{equation}
\mathcal{L}_{a}\rightarrow \QTR{sl}{LL}_{a}.
\end{equation}
This idea has been used in \cite{14} to study Haldane hierarchical states.
We will skip the details here and goes to the second realization which
interest us in this paper.

The second scenario, for studding single layer hierarchical states, at level
$k$, is based on the physical idea that, instead\ of having only the first
Landau level occupied like in Laughlin model; one can have as well other
neighboring Landau levels occupied by electrons. This is mainly the content
of the idea behind hierarchical models involving several kinds of
quasi-particles like in Haldane case, eq(1.2); see figures 2.

\begin{figure}[tbh]
\vspace{1.5cm}
\begin{equation*}
\epsfxsize=7cm \epsffile{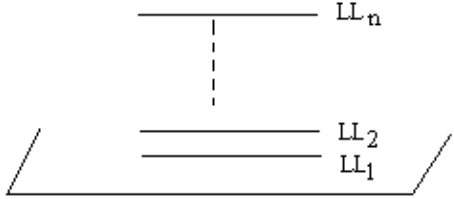}
\end{equation*}
\caption{\textit{This figure illustrates a single layer} $\mathcal{L}$
\textit{\ where the first }$n$\textit{\ Landau levels are occupied. We
suppose that all levels are of Laughlin type as in the decompositions
eqs(1.1-3). The way in which these Landau levels are occupied depends on the
model one is considering. In general one may have different filling factors
as in eq(1.2) and possible interactions between the various Levels. }}
\label{figure 2}
\end{figure}
This scheme may be also be interpreted as describing $k$ branches of FQH
liquid as in the case of hydrodynamic droplets and edge excitations of
systems with boundaries \cite{21}. From the mathematical view, both of the
bulk and edge levels can be implemented in terms of fiber bundles idea; see
figures 3.
\begin{figure}[tbh]
\vspace{1.5cm}
\begin{equation*}
\epsfxsize=8.5cm \epsffile{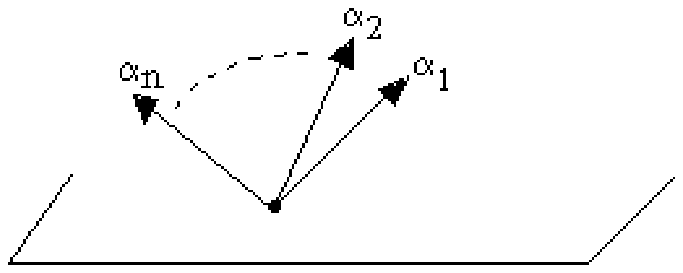}
\end{equation*}
\caption{\textit{This figure represents a D2 brane and n dimensional fiber
above a point on D2. Each direction} $\protect\alpha _{i},i=1,...,n;$
\textit{\ of this fiber is associated with the i-th Landau level of the FQH
layer realized here by the D2 brane. From this view point, Laughlin model
with filling factor }$1/p$\textit{\ is then associated with a one
dimensional fiber bundle generated by the } $\protect\alpha _{1}$ {\ vector
with a norm equal to the odd integer p.}}
\end{figure}
In the fiber bundle picture, one insists on having the correspondence (\ref
{cor1}); while the $i-th$ Landau Level $\QTR{sl}{LL}_{i}$ is thought of as a
fiber \textsl{F}$^{i}$ above the $\mathcal{D}2$ brane which now plays the
role of a base manifold $B$;
\begin{equation}
\left( \QTR{sl}{LL}_{k};\mathcal{L}\right) \longleftrightarrow \QTR{sl}{F}%
^{k}\left( {\large B}\right) \longleftrightarrow \left( \QTR{sl}{V}_{k}%
\QTR{sl}{;}\mathcal{D}2\right) ,  \label{cor2}
\end{equation}
where $\QTR{sl}{V}_{k}$\ is a $k$ dimension vector space on the set of
integers and is defined globally on the base. In this way, one has at hand
various $U\left( 1\right) $ gauge fields components $A_{\mu }^{i}\left(
y\right) $; i.e,
\begin{equation}
A_{\mu }\left( \QTR{sl}{LL}_{i};\mathcal{L}\right) \longleftrightarrow
A_{\mu }^{i}\left( \mathcal{L}\right) ,  \label{cor3}
\end{equation}
to describe the various $\left\{ \QTR{sl}{LL}_{i}\right\} $ Landau levels%
\footnote{%
The fiber bundle we are using here has formal similarities with the one used
in the construction of conformal and affine Toda theories \cite{a}. The
unique difference is that in Toda models the $\alpha _{i}$'s are the roots
of a Lie algebra; say $U\left( n\right) $, and $\alpha _{i}.\alpha _{j}$ is
the $U\left( n\right) $ Cartan matrix. In this paper, the $\alpha _{i}$'s
are general objects as they depend on the filling factors and have quantized
norms.}. The effective field theory one expects is a Chern Simons gauge
field model on the fiber bundle \textsl{F}$\left( B\right) $, which we want
to build now.

The single layer $\mathcal{L}$, viewed now as a $\mathcal{D}2$ brane, has
world volume parameterized by the $\left\{ y=\left( y^{\mu }\right) \right\}
$ coordinates. Fiber bundles $\QTR{sl}{F}\left( \mathcal{L}\right) $ based
on $\mathcal{L}$\ are generally given by the union of \ fibers $\QTR{sl}{F}%
_{y}$ based on each point $y$ of the base.
\begin{equation}
\QTR{sl}{F}\left( \mathcal{L}\right) =\cup _{y}\QTR{sl}{F}_{y}\left(
\mathcal{L}\right)
\end{equation}
Each fiber $\QTR{sl}{F}_{y}$\ is a vector space parameterized by $\left(
\alpha _{i};y\right) $, where $\alpha _{i}$ are some linearly independent
vectors defining the vector basis of $\QTR{sl}{F}_{y}$. Since in our present
case, the $\alpha _{i}$'s are $y$\ independent, the fibers $\QTR{sl}{F}_{y}$
are the same every where on the base; this is why we shall drop the $y$
index on $\QTR{sl}{F}_{y}$ which now on will be denoted as $\QTR{sl}{V}%
\left( \mathcal{L}\right) $. Moreover since the level $k$ of the hierarchy
can be any positive integer, $\QTR{sl}{V}\left( \mathcal{L}\right) $ should
be a priori an infinite dimensional vector space which we endow with the
orthonormal canonical basis.
\begin{equation}
\left\{ \mathbf{e}_{i};\quad \mathbf{e}_{i}\cdot \mathbf{e}_{i}=\delta
_{ij};\quad i,j\in \mathbf{Z}\right\} .
\end{equation}
Though we will deal only with finite $m$ dimension proper subspaces $%
\QTR{sl}{V}_{m}\left( \mathcal{L}\right) $ of $\QTR{sl}{V}\left( \mathcal{L}%
\right) $ since generally one is interested to the few leading terms of the
hierarchy; one can address the mathematical structure of $\QTR{sl}{V}\left(
\mathcal{L}\right) $ in its general form. Problem induced by the infinite
series one my encounters, such as the convergence of trace on infinite
matrices for instance, may be regularized as in \cite{raina,24}.
Furthermore, as hierarchy at generic levels is described in a recurrent
manner, namely
\begin{equation}
\text{level }1\subset \text{level }2\subset ...\subset \text{level }\left(
n-1\right) \subset \text{level }n\subset ...
\end{equation}
we demand that the proper subspaces $\QTR{sl}{V}_{n}\left( \mathcal{L}%
\right) $ of $\QTR{sl}{V}\left( \mathcal{L}\right) $ have to verify,
\begin{equation}
\QTR{sl}{V}_{1}\left( \mathcal{L}\right) \subset \QTR{sl}{V}_{2}\left(
\mathcal{L}\right) \subset ...\subset \QTR{sl}{V}_{n}\left( \mathcal{L}%
\right) \subset ...\subset \QTR{sl}{V}\left( \mathcal{L}\right) .
\end{equation}
Note that a generic proper subspace $\QTR{sl}{V}_{m}\left( \mathcal{L}%
\right) $ has a canonical basis induced from the mother space $\QTR{sl}{V}%
\left( \mathcal{L}\right) $. However, because of the physics we are looking
to describe, we will introduce the special $\left\{ \alpha _{i};1\leq i\leq
m\right\} $ vectors basis of $\QTR{sl}{V}_{m}\left( \mathcal{L}\right) $
with the property of having non zero intersection matrix; that is the $%
\alpha _{i}$'s form a non orthogonal basis taken as,
\begin{equation}
\alpha _{i}\cdot \alpha _{j}=\mathrm{G}_{ij}  \label{wenmatr1}
\end{equation}
The matrix $\mathrm{G}_{ij}$ appearing in this relation is a hermitian and
invertible $m\times m$ matrix; it will be interpreted as the matrix
appearing in the Wen-Zee effective fields model \cite{28,281,29,291,292,30}.
So we shall refer to this basis as Wen-Zee one and then to the fiber bundle
as Wen-Zee fiber. We will also see later on, when we study NC effective
field models for hierarchical states going beyond the NC Susskind model;
that it is useful to extend the sequence (2.14) by introducing a one
dimension bundle $\QTR{sl}{V}_{0}\left( \mathcal{L}\right) $ associated with
the constant background appearing in the Susskind map that led to the
discovery of NC Chern Simons model for Laughlin states. As such, we have
\begin{equation}
\QTR{sl}{V}_{0}\left( \mathcal{L}\right) \subset \QTR{sl}{V}_{1}^{\prime
}\left( \mathcal{L}\right) \subset \QTR{sl}{V}_{2}^{\prime }\left( \mathcal{L%
}\right) \subset ...\subset \QTR{sl}{V}_{n}^{\prime }\left( \mathcal{L}%
\right) \subset ...\subset \QTR{sl}{V}\left( \mathcal{L}\right) ,
\end{equation}
where now $\QTR{sl}{V}_{m}^{\prime }\left( \mathcal{L}\right) $\ have $%
\left( m+1\right) $\ dimensions. The new vector basis is then $\left\{
\alpha _{i};0\leq i\leq m\right\} $ and the previous $\mathrm{G}_{ij}$
intersection matrix now becomes a $\left( m+1\right) \times \left(
m+1\right) $ matrix.
\begin{equation}
\mathrm{S}_{ij}=\left(
\begin{array}{cc}
\mathrm{G}_{00} & \mathrm{G}_{0j} \\
\mathrm{G}_{i0} & \mathrm{G}_{ij}
\end{array}
\right)  \label{wenmatr2}
\end{equation}
Besides the useful choice $\mathrm{S}_{00}=1$ we will make later, we require
that this generalized matrix is hermitian and invertible as well.

\subsubsection{Case of Multilayers}

Hierarchical states of multilayer system are described by a general fiber
bundle $\left( \QTR{sl}{V};B\right) $ whose base {\large B}$=\left\{ n%
\mathcal{D}2\right\} $ and fiber \textsl{V}$=\left\{ \cup _{a=1}^{n}\cup
_{i_{a}=1}^{m_{a}}\QTR{sl}{LL}_{i_{a}}\right\} $. The positive integer
indices $m_{a}$\ define the levels of hierarchy in each layer $\mathcal{L}%
_{a}$; while $n$ is the number of parallel layers.
\begin{eqnarray}
\left\{ \cup _{a=1}^{n}\mathcal{L}_{a}\right\} &\longleftrightarrow &n%
\mathcal{D}2\longleftrightarrow \text{ base {\large B}},  \notag \\
\left\{ \cup _{a=1}^{n}\cup _{i_{a}=1}^{m_{a}}\QTR{sl}{LL}_{i_{a}}\right\}
&\longleftrightarrow &\left( \text{\textsl{V;}}n\mathcal{D}2\right)
\longleftrightarrow \text{ fiber \textsl{V }over\textsl{\ \ }{\large B}}
\end{eqnarray}
In this correspondence, the gauge fields components $A_{\mu }\left( y\right)
$, describing the displacement of the hierarchical Hall fluid, carry a non
abelian $U\left( n\right) $ structure, inherited from the symmetry of the
base $B$, and behave as a vector as in (\ref{cor3}); i.e,
\begin{equation}
A_{\mu }\left[ \cup _{a=1}^{n}\left( \cup _{i_{a}=1}^{m_{a}}\QTR{sl}{LL}%
_{i_{a}};\mathcal{L}_{a}\right) \right] \longleftrightarrow \left( A_{\mu
}\right) _{ab}^{i};\quad a,b,1,...,n;\quad i=1,...,m\text{,}
\end{equation}
where $m=\left( m_{1}+...+m_{n}\right) $. Note that the $a$ and $b$ indices
carried by this gauge fields refer to the interaction between electrons
moving in layers $\mathcal{L}_{a}$ and $\mathcal{L}_{b}$\ while the index $i$%
\ encodes interactions between electrons sitting in Landau levels of the
whole FQH system. Here also there is an analogue of the intersection
matrices (\ref{wenmatr1}) and (\ref{wenmatr2}); the only difference is that
now the dimension $m$ is given by the sum over the $m_{a}$ dimensions. For
practical reasons, we shall often use the canonical decomposition of the $%
U\left( n\right) $ hermitian gauge fields $\mathbf{A}_{\mu }$ of this system
as,
\begin{equation}
\mathbf{A}_{\mu }=\sum_{a,b=1}^{n}\mathbf{A}_{\mu }^{ab}\text{\textsl{u}}%
_{ab};\quad \text{\textsl{u}}_{ab}=|a><b|  \label{adj1}
\end{equation}
Since this gauge field is also a vector under hierarchy, one can also
expanded it as a linear combination using the $\alpha _{i}$\ vector basis
as,
\begin{equation}
\mathbf{A}_{\mu }=\sum_{i=1}^{m_{1}+...+m_{n}}A_{\mu }^{i}\text{ }\alpha _{i}
\label{hier1}
\end{equation}
Each component $\mathbf{A}_{\mu }^{ab}$ of the expansion (\ref{adj1}) of the
matrix gauge field $\mathbf{A}_{\mu }$\ has as well a decomposition of the
type (\ref{hier1}).
\begin{figure}[tbh]
\vspace{1.5cm}
\begin{equation*}
\epsfxsize=8.5cm \epsffile{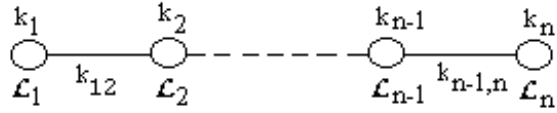}
\end{equation*}
\caption{\textit{This is a geometric representation of a multilayer FQH
state. Each a-th node, of the diagram} $\left\{ \cup _{b=1}^{n}{\mathcal{L}}%
_{b}\right\} $, \textit{\ describes a layer }${\mathcal{L}}_{a}$\textit{;
viewed as a spherical }$D2$\textit{\ brane with filling factor }$k_{a}$.
\textit{\ Nodes are associated with the a-th projector} $\protect\pi _{a}$
\textit{\ on the }$n$\textit{\ dimensional basis vector } $|a>$ \textit{\
and the filling factor} $k_{a}$. \textit{\ Interactions between layers are
restricted to the closest ones }$\left( \mathcal{L}_{a},\mathcal{L}_{a\pm
1}\right) $\textit{\ and are encoded in the links }$u_{ab}=|a><b|$\textit{\
with integral coefficients }$k_{ab}$. \textit{\ Such diagram, which is of
the ordinary } $A_{n-1}$ \textit{\ type classification, has a formal
similarity with the geometric engineering of the Coulomb branch of N=2
supersymmetric theories \protect\cite{A,B}.}}
\label{figure 4}
\end{figure}
Having fixed our terminology and convention notations, we now turn to study
the effective field models for multilayer states occupying the first Landau
level of the system; but also those general states occupying several Landau
levels.

\section{FQH Effective Field Model}

\qquad As we have several situations depending on the number of layers as
well as the number of Landau levels on the layers represented by the fiber
bundles $\left\{ \cup _{a=1}^{n}\left\{ \cup _{i_{a}=1}^{m_{a}}\QTR{sl}{LL}%
_{i_{a}};\mathcal{L}_{a}\right\} \right\} $, one may distinguish different
kinds of interactions coming from:

(i) the internal structure of the base manifold {\large B}$=\cup _{a=1}^{n}%
\mathcal{L}_{a}$, which in the case of coincident layers lead a priori non
abelian symmetries generated by $\left\{ |a><b|\right\} $ as in we usually
have in Brane systems.

(ii) couplings of the fibers \textsl{F=}$\cup _{i_{a}=1}^{m_{a}}\QTR{sl}{LL}%
_{i_{a}}$ living on the base manifold. In the case where all layers are in
the Laughlin ground states for instance; that is a fiber bundle of the form $%
\cup _{a=1}^{n}\left\{ \QTR{sl}{LL}_{1};\mathcal{L}_{a}\right\} $; fbers
interactions are encoded in the intersection matix $\alpha _{a}.\alpha _{b}=%
\mathrm{G}_{ab}$.

(iii) both from the  base {\large B} and the fiber \textsl{F}.

We will then adopt the following strategy in studying the effective field
model for the multilayer systems. First we suppose that all layers of the
system are of Laughlin type and describe the commutative field approach.
Then, we study its non commutative extension using Susskind method. Once
this done, we consider the case a single layer; but with several Landau
levels. Here also we study first the commutative field model; and then we
analyze its non commutative extension. Finally, we consider the case of
several layers and several Landau levels.

\subsection{Multilayer Commutative Field Model}

\qquad To get the generalized effective field model describing multilayer
FQH states, one should specify the condition on $\left| d_{a}-d_{b}\right| $
spacings. The point is that the $A_{\mu }$ gauge fields on the multilayers
have, in addition to the $n\times n$ matrix structure, a dependence on the $%
y_{a}^{\mu }=\left( y^{1},y^{2},y^{3}=d_{a}\right) $ coordinates of the $n$
layers, viewed as parallel $D2$ branes. According to the values of $\left|
d_{a}-d_{b}\right| $, one distinguishes several cases lying between the
completely abelian model with a $U\left( 1\right) ^{n}$ gauge symmetry and
the largest non abelian $U\left( n\right) $ invariance.

\subsubsection{Abelian model}

\qquad If for example, one supposes that $\left| d_{a}-d_{b}\right|
>>>1,\quad \forall a,b=1,...,n$; that is all layers are very distant from
each others, then layers interactions may be ignored and one is left with a $%
U\left( 1\right) ^{n}$\ model.
\begin{figure}[tbh]
\vspace{1.5cm}
\begin{equation*}
\epsfxsize=6cm \epsffile{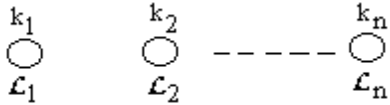}
\end{equation*}
\caption{\textit{This figure represents a system of }$\mathit{n}$\textit{\
parallel layers very distant from each others. They describe }$n$\textit{\
uncoupled FQH states with filling factor} $\protect\nu _{a}=1/k_{a}$.\textit{%
\ The gauge field are given by the diagonal terms and the gauge symmetry is }
$U(1)^{n}$. \textit{The total filling factor of the system is just the sum.}}
\label{figure 5}
\end{figure}
In this system, each one of the layers ( $D2$ brane ), partially filled with
$\nu _{a}$ of the form $\frac{1}{k_{a}}$, $k_{a}$ odd integer, describes an
electronic system with a conserved number $\mathcal{N}_{a}$ of particles%
\footnote{%
Since $\mathrm{\nu }=\frac{\mathcal{N}_{e}}{\mathcal{N}_{\phi }}=$ constant
and as the external magnetic field $\mathcal{B}_{ex}$ is taken constant, the
number of quantum flux $\mathcal{N}_{\phi }$\ $=$ $\int_{\mathcal{L}}%
\mathcal{B}_{ex}d^{2}\sigma =\mathcal{B}_{ex}Surface\left( \mathcal{L}%
\right) $ \ \ is constant provided the surface $Surface\left( \mathcal{L}%
\right) $ of the layer $\mathcal{L}$ does. In this case the number $\mathcal{%
N}_{e}$ of particles can be viewed as a constant of motion.},
\begin{equation}
\frac{d\mathcal{N}_{a}}{dt}=0,
\end{equation}
In effective field theory, the number $\mathcal{N}_{a}$ of electrons moving
on the layer may be also expressed as an integral over the density electron
number par surface unit;
\begin{equation}
\mathcal{N}_{a}=\int_{\mathcal{L}_{a}}d^{2}y\text{ }\mathcal{J}%
_{a}^{0}\left( y,d_{a}\right) ,
\end{equation}
where $\mathcal{J}_{a}^{0}\left( y,d_{a}\right) =\partial ^{2}\mathcal{N}%
_{a}/\partial ^{2}y$. The condition of conservation (3.1) is ensured by the
introduction of a $\left( 1+2\right) $ conserved current $\mathcal{J}%
_{a}^{\mu }$; i. e $\partial _{\mu }\mathcal{J}_{a}^{\mu }=\partial _{0}%
\mathcal{J}_{a}^{0}+\partial _{i}\mathcal{J}_{a}^{i}=0$ leading to\
\begin{eqnarray}
\frac{d\mathcal{N}_{a}}{dt} &=&-\int_{\mathcal{L}_{a}}d^{2}y\partial _{i}%
\mathcal{J}_{a}^{i}  \label{cur1} \\
&=&-\frac{1}{2\pi }\oint_{\partial \mathcal{L}_{a}}dl_{i}A_{i}^{a},
\end{eqnarray}
which vanishes provided the boundary $\partial \mathcal{L}_{a}$ of the layer
is zero, or the $A_{a}^{i}$ field vanishes on $\partial \mathcal{L}_{a}$ or
again due to the existence of some periodicity features. The equality of
eq(3.4), using the hypothesis $\mathcal{J}_{a}^{i}=\frac{1}{2\pi }%
\varepsilon ^{0ij}\partial _{i}A_{j}^{a}$, reflects just the Chern Simons
field realization of the $\mathcal{J}_{a}^{\mu }$ conserved current, which
reads as,
\begin{equation}
\mathcal{J}_{a}^{\mu }\left( y,d_{a}\right) =\frac{1}{2\pi }\varepsilon
^{\mu \nu \sigma }\partial _{\nu }A_{\sigma }^{a};\quad a=1,...,n,
\end{equation}
Since the layers $\mathcal{L}_{a}$ are partially filled with filling factor$%
1/\mathrm{k}_{a}$, we should have the remarkable relations,
\begin{equation}
\mathrm{k}_{a}\mathcal{N}_{a}=\mathcal{N}_{\phi };\quad a=1,...,n,
\end{equation}
which, in the framework of the Chern Simons model, this is just the equation
of motion of the time component of the gauge fields; that is
\begin{equation}
\mathrm{k}_{a}\frac{1}{2\pi }\varepsilon ^{0\nu \sigma }\partial _{\nu
}A_{\sigma }^{a}=\frac{1}{2\pi }\varepsilon ^{0\nu \sigma }\partial _{\nu }%
\mathcal{A}_{\sigma },
\end{equation}
without summation on the repeated index $a$. The original formal action,
invariant under the gauge changes $\delta A_{\mu }^{a}=\partial _{\mu
}\lambda ^{a}$, and leading to the above eq of motion is,
\begin{eqnarray}
S\left[ A_{\mu }^{a},\mathcal{A}_{\mu }\right]  &=&\frac{\mathrm{1}}{4\pi }%
\int d^{3}y\varepsilon ^{\mu \nu \sigma }\sum_{a=1}^{n}\mathrm{k}_{a}A_{\mu
}^{a}\partial _{\nu }A_{\sigma }^{a}  \notag \\
&&-\frac{\mathrm{1}}{2\pi }\int d^{3}y\varepsilon ^{\mu \nu \sigma }\left(
\sum_{a=1}^{n}A_{\mu }^{a}\right) \partial _{\nu }\mathcal{A}_{\sigma }.
\end{eqnarray}
Note that the right way one should think about this action is in terms of
the field diagonal matrices,
\begin{equation}
A_{\mu }=\sum_{a=1}^{n}A_{\mu }^{a}\pi _{a};\quad \pi _{a}=|a><a|,
\end{equation}
where $\pi _{a}$'s are the projectors on the layers. In terms of these
matrices, the previous action may be rewritten in a more interesting form
as,
\begin{equation}
S\left[ A_{\mu }^{a},\mathcal{A}_{\mu }\right] =\frac{\mathrm{1}}{4\pi }\int
d^{3}y\varepsilon ^{\mu \nu \sigma }\left[ Tr\left( \mathrm{\kappa }A_{\mu
}\partial _{\nu }A_{\sigma }-2A_{\mu }\partial _{\nu }\mathcal{A}_{\sigma
}\right) \right] ,  \label{actionm}
\end{equation}
Here $\mathrm{\kappa }$ is the coupling matrix, itself a $n\times n$
diagonal\ matrix,
\begin{equation}
\mathrm{\kappa =}\sum_{a=1}^{n}\mathrm{k}_{a}\pi _{a}.
\end{equation}
By help of the $A_{\sigma }$ and $\mathrm{\kappa }$ matrices, the new eq of
motion following from the action(\ref{actionm}) is,
\begin{equation}
\mathrm{\kappa }\frac{1}{2\pi }\varepsilon ^{0\nu \sigma }\partial _{\nu
}A_{\sigma }=\frac{1}{2\pi }\varepsilon ^{0\nu \sigma }\partial _{\nu }%
\mathcal{A}_{\sigma },
\end{equation}
which, upon integration over $d^{2}y$, gives the $n\times n$ matrix eq
\begin{equation}
\mathrm{\kappa }\mathcal{N}\mathrm{=}\mathcal{N}_{\phi }I;\quad \quad
\mathcal{N}\mathrm{=}\mathcal{N}_{a}\pi _{a}
\end{equation}
Taking the trace over this relation, one gets the total filling factor
\begin{eqnarray}
\mathrm{\nu } &=&Tr\left( \mathrm{\kappa }^{-1}\right)   \label{formajic} \\
&=&\sum_{a=1}^{n}\frac{1}{\mathrm{k}_{a}}.  \notag
\end{eqnarray}
Remember the first term of this relation; we shall see that it is also valid
in presence of interactions and whether the gauge symmetry is abelian or
not. Note that the appearance of the $\mathrm{\kappa }$\ coupling matrix
seems a little bit strange. In fact this is a very special point inherent to
the $U\left( 1\right) ^{n}$ abelian symmetry generated by the projectors $%
\pi _{a}$. In the non abelian case, the coupling should, under some
conditions to be specified in subsubsection 3.1.2 when studying the second
kind of solutions, be an integer. Before that let us first derive the
appropriate constraint eqs linking gauge symmetry and filling factors and
then return to complete the comment regarding eq(\ref{formajic}).

\subsubsection{Non abelian Case}

\qquad Non abelian FQH systems appear in presence of interactions between
the $\mathcal{L}_{a}$ layers of the system $\left\{ \cup _{a=1}^{n}\mathcal{L%
}_{a}\right\} $, as in the special case where all layers are quasi
coincident; i.e all parameters $d_{a}\sim d_{1}$; i.e
\begin{equation}
\left| d_{a}-d_{b}\right| <<<1,\quad \forall a,b=1,...,n,
\end{equation}
or more generally as in cases where subsets $\left\{ \cup _{a_{s}=1}^{r}%
\mathcal{L}_{a_{s}};r<n\right\} $ of the layers are closed enough to each
others while the remaining others $\left\{ \cup _{a_{s}=r+1}^{n}\mathcal{L}%
_{a_{s}};\right\} $ are far away;
\begin{equation}
\left| d_{a_{s}}-d_{b_{s}}\right| <<<1,\quad a_{s},b_{s}\in J_{r}\subset
\left[ 1,n\right] .
\end{equation}
In this case, the $N_{a}$ numbers are no longer constants of motion since
particles can travel from a given layer $\mathcal{L}_{a}$ to an other closed
$\mathcal{L}_{b}$ one. To illustrate this feature more explicitly, let $%
\mathcal{N}$ denote the total number of electrons moving in the multilayer
system,
\begin{equation}
\mathcal{N}=\sum_{a=1}^{n}N_{a}.
\end{equation}
Because of interaction between layers, this number can usually written as
the trace over a $n\times n$\ matrix $\mathbf{N=}\left( \mathbf{N}%
_{ab}\right) $, in complete harmony with the previous case where
interactions were ignored,
\begin{equation}
\mathcal{N}=Tr\mathbf{N=}\sum_{a=1}^{n}N_{a}  \label{totnb}
\end{equation}
In fact the effective number $\widetilde{N}_{a}$ of electrons moving on a
given layer $\mathcal{L}_{a}$, at a given time, is the sum of three terms;
namely the initial number $N_{a}$ \textit{plus} the number $N_{ab}$ of
electrons leaving the $\left( n-1\right) $ layers $\mathcal{L}_{b}$ of the
system and landing on $\mathcal{L}_{a}$ \textit{minus} the number $N_{ba}$
of electrons leaving $\mathcal{L}_{a}$ for the other $\mathcal{L}_{b}$
layers,
\begin{equation}
\widetilde{N}_{a}=N_{a}+\sum_{b=1,b\neq a}^{n}N_{ab}-\sum_{b=1,b\neq
a}^{n}N_{ba};\quad a=1,...,n,
\end{equation}
This relation may be also rewritten, by adding and subtracting the numbers $%
N_{a}\equiv N_{aa}$, as follows
\begin{equation}
\widetilde{N}_{a}=N_{a}+\sum_{b=1}^{n}\left( N_{ab}-N_{ba}\right) ;\quad
a=1,...,n,
\end{equation}
Summing over all $\widetilde{N}_{a}$'s, one discovers the following
conserved quantity,
\begin{equation}
\mathcal{N}=\sum_{a=1}^{n}\widetilde{N}_{a}=\sum_{a=1}^{n}N_{a}
\end{equation}
To get the effective field model describing the multilayer FQH states with
interactions; we shall follow the method we have developed above by
expressing the constant of motion $\mathcal{N}$ as an integral over a
density $\mathcal{\rho }$ of a conserved current $\mathcal{J}^{\mu }$; that
is
\begin{eqnarray}
\mathcal{N} &=&\int_{\cup \mathcal{L}_{a}}d^{2}y\text{ }\mathcal{\rho }; \\
0 &=&\partial _{0}\mathcal{\rho }+\partial _{i}\mathcal{J}^{i}.
\end{eqnarray}
Since the FQH system $\left\{ \cup _{a=1}^{n}\mathcal{L}_{a}\right\} $ has
interactions one expects, from the similarity between $\left\{ \cup
_{a=1}^{n}\mathcal{L}_{a}\right\} $ and the assembly of $n$ parallel $D2$
branes, that the underlying theory is, roughly speaking, a non abelian Chern
Simons $U\left( n\right) $ gauge model. This is indeed what one gets if one
does not worry about the details of FQH physics on each layer and insists on
$U\left( n\right) $ gauge symmetry. Indeed having $U\left( n\right) $ gauge
invariance requires coincident $D2$ branes but moreover the same filling
factor. To get the point let us set the problem in its general form and work
out explicitly the link between gauge invariance and layers filling factors.

\begin{itemize}
\item  \textsl{Constraint Eqs}
\end{itemize}

Since the total number $\mathcal{N}$ of electrons is a constant of motion
following from the existence of a Noether conserved current $\mathcal{J}%
^{\mu }$ and since $\mathcal{N}$\ has the form of a trace (\ref{totnb}), let
us write $\mathcal{J}^{\mu }$ as
\begin{equation}
\mathcal{J}^{\mu }=\mathrm{Tr}\mathbf{J}^{\mu },
\end{equation}
where $\mathbf{J}^{\mu }$ is a $n\times n$ hermitian\ matrix. Moreover,
since it is the trace $\mathrm{Tr}\mathbf{J}^{\mu }$ that should be
conserved and not necessary the matrix $\mathbf{J}^{\mu }$; one may in
general realize this matrix field in terms of the non abelian Chern Simons
potential as,
\begin{equation}
\mathbf{J}^{\mu }=\frac{1}{2\pi }\varepsilon ^{\mu \nu \sigma }\left(
\mathbf{P.}\partial _{\nu }\mathbf{A}_{\sigma }+i\mathbf{Q.}\left[ \mathbf{A}%
_{\nu },\mathbf{A}_{\sigma }\right] \right) ,
\end{equation}
where the couplings $\mathbf{P}$ and $\mathbf{Q}$ are a priori arbitrary
invertible and hermitian $n\times n$ matrices with integer entries; that is $%
\mathbf{P,Q\in }Gl\left( n,Z\right) $. Although the idea of taking $\mathbf{P%
}$ and $\mathbf{Q}$ couplings as matrices\ seems going against what is
established in non abelian Chern Simons gauge theory; it is however a
physical argument extending eq(3.11) which allows us to get the constraint
eqs between gauge invariance and the filling factors of the layers. So $%
\mathbf{P}$ and $\mathbf{Q}$ as matrices is dictated by the fact that we
want to insert in the effective gauge field model, we looking to build, the
fact that layers may have different filling factors. Requiring non abelian
gauge invariance, which acts as,
\begin{eqnarray}
\mathbf{A}^{\prime } &=&\mathbf{U}^{-1}\left( \mathbf{A}-\partial \right)
\mathbf{U};  \notag \\
\mathbf{J}^{\mu \prime } &=&\mathbf{U}^{-1}\mathbf{J}^{\mu }\mathbf{U,}
\end{eqnarray}
where $\mathbf{U}$\ is a $n\times n$\ matrix defining the gauge
transformations as well as the condition of current conservation $\partial
_{\mu }\mathcal{J}^{\mu }=0$, we get the following constraint eqs,
\begin{eqnarray}
\mathbf{U}^{-1}\mathbf{PU} &=&\mathbf{P,} \\
\mathbf{U^{-1}QU} &=&\mathbf{Q,}  \label{contr1}
\end{eqnarray}
and
\begin{equation}
\frac{1}{2\pi }\varepsilon ^{\mu \nu \sigma }\partial _{\mu }\mathrm{Tr}%
\left( \mathbf{P.}\partial _{\nu }\mathbf{A}_{\sigma }+i\mathbf{Q.}\left[
\mathbf{A}_{\nu },\mathbf{A}_{\sigma }\right] \right) =0,  \label{constr2}
\end{equation}
The two first constraint eqs (3.27,\ref{contr1}) reflect the fact that $%
\mathbf{P}$ and $\mathbf{Q}$ should be gauge invariant in order to be
interpreted as physical coupling constants; while the third one (\ref
{constr2}) implies that the total number of particles is a constant of
motion. Since $\mathbf{U}$ is an arbitrary\ gauge transformation matrix,
these constraint eqs impose a severe restriction on the kind of $\mathbf{P}$
and $\mathbf{Q}$ matrices one can have. Let us explore the possible
solutions on illustrating examples.

\begin{itemize}
\item  \textsl{Solutions}
\end{itemize}

(1) \textit{Solution I: Non Abelian }$U\left( n\right) $\textit{\ Chern
Simons model}

If we insist on having $U\left( n\right) $ gauge invariance, that is $%
\mathbf{U}$ an arbitrary $n\times n$ unitary matrix, then there is a unique
solution that commute with all $U\left( n\right) $ matrix elements. This
solution is proportional to the identity as required by the Schur lemma;
that is a number $\mathrm{k}$ times the identity operator
\begin{equation}
\mathbf{P=}\mathrm{k}^{\prime }\text{ }\mathbf{\mathrm{I;\quad }Q=}\mathrm{k}%
^{\prime \prime }\text{\textrm{\ }}\mathbf{\mathrm{I}}\text{,}
\end{equation}
Unitarity require moreover that \textrm{k=}$\mathrm{k}^{\prime }/\mathrm{k}%
^{\prime \prime }$ should be positive integer. In this case the solution for
$\mathbf{J}^{\mu }$ in terms of the $U\left( n\right) $ gauge potentials $%
\mathbf{A}_{\mu }$ reads, after setting $\mathrm{k}^{\prime \prime }=1$ or
absorbing it in the gauge field ( $\mathbf{A\rightarrow }\mathrm{k}^{\prime }%
\mathbf{A}$ ), as
\begin{eqnarray}
\mathbf{J}^{\mu } &=&\frac{\mathrm{k}}{2\pi }\varepsilon ^{\mu \nu \sigma
}\left( \partial _{\nu }\mathbf{A}_{\sigma }+i\left[ \mathbf{A}_{\nu },%
\mathbf{A}_{\sigma }\right] \right) , \\
\mathbf{J}^{0} &=&\frac{\mathrm{k}}{2\pi }\varepsilon ^{0\nu \sigma }\left(
\partial _{\nu }\mathbf{A}_{\sigma }+i\left[ \mathbf{A}_{\nu },\mathbf{A}%
_{\sigma }\right] \right)
\end{eqnarray}
Note that though non abelian, the $\mathbf{J}^{0}$ time component may also
be put in the useful form,
\begin{equation}
\mathbf{J}^{0}=\frac{\mathrm{k}}{2\pi }\mathbf{B}
\end{equation}
where now the $n\times n$ matrix $\mathbf{B}$ is the curl of the non abelian
gauge potential field $\mathbf{A}_{\mu }$. Observe in passing that while $%
\mathcal{J}^{\mu }=\mathrm{Tr}\mathbf{J}^{\mu }$ is conserved due to the
obvious property $\mathrm{Tr}\left[ \mathbf{A}_{\nu },\mathbf{A}_{\sigma }%
\right] =0$, the $\mathbf{J}^{\mu }$ vector field is not conserved in the
usual sense since in general $\partial _{\mu }\left( \varepsilon ^{\mu \nu
\sigma }\left[ \mathbf{A}_{\nu },\mathbf{A}_{\sigma }\right] \right) $ is
different from zero. This property reflects the fact that the numbers $N_{a}$
of particles on \ each layer $\mathcal{L}_{a}$ are not constants of motion
in the usual sense. The $\mathbf{J}^{\mu }$ current obeys however a
covariantly conserved relation namely,
\begin{equation}
D_{\mu }\mathbf{J}^{\mu }=\partial _{\mu }\mathbf{J}^{\mu }-i\left[ \mathbf{A%
}_{\mu },\mathbf{J}^{\mu }\right] =0.
\end{equation}
Moreover, from the identity $\mathcal{N}=\nu \mathcal{N}_{\phi }$, or again
by equating the current densities, we have,
\begin{equation}
\frac{\mathrm{k}}{2\pi }\mathbf{B}=\frac{\mathrm{1}}{2\pi }\mathcal{B}_{ex}.%
\mathrm{I}
\end{equation}
This relation reads in terms of the non abelian $\mathbf{A}$ gauge fields
and the $\varepsilon ^{0\nu \sigma }$ antisymmetric tensor as,
\begin{equation}
\frac{\mathrm{k}}{2\pi }\varepsilon ^{0\nu \sigma }\left( \partial _{\nu }%
\mathbf{A}_{\sigma }+i\left[ \mathbf{A}_{\nu },\mathbf{A}_{\sigma }\right]
\right) =\frac{\mathrm{1}}{2\pi }\varepsilon ^{0\nu \sigma }\partial _{\nu }%
\mathcal{A}_{\sigma }.\mathrm{I,}
\end{equation}
This way of writing shows that one is dealing just with the equation of
motion of the time component $\mathbf{A}_{0}$ of the gauge field $\mathbf{A}%
_{\mu }$. This eq together with the two remaining others recovered under $%
SO\left( 1,2\right) $ covariance, follow from the action the non abelian $%
U\left( n\right) $ Chern Simons gauge field theory,
\begin{eqnarray}
S\left[ A_{\mu },\mathcal{A}_{\mu }\right] &=&\frac{\mathrm{k}}{4\pi }\int
d^{3}y.\varepsilon ^{\mu \nu \sigma }\mathrm{Tr}\left( \mathbf{A}_{\mu
}\partial _{\nu }\mathbf{A}_{\sigma }+\frac{2i}{3}\mathbf{A}_{\mu }\left[
\mathbf{A}_{\nu },\mathbf{A}_{\sigma }\right] \right)  \notag \\
&&-\frac{\mathrm{1}}{2\pi }\int d^{3}y.\varepsilon ^{\mu \nu \sigma }\mathrm{%
Tr}\left( \mathbf{A}_{\mu }\partial _{\nu }\mathcal{A}_{\sigma }\right)
\end{eqnarray}
The filling fraction of the multilayer system is
\begin{equation}
\mathrm{\nu }=\frac{1}{\mathrm{k}}\mathrm{Tr}\left( \mathbf{I}\right)
\mathrm{=}\frac{n}{\mathrm{k}}.
\end{equation}
This result, which was expected from general features of non abelian Chern
Simons gauge theory and branes physics, may be exploited to derive the
general expression of the filling factors \ associated with FQH multilayer
states having a gauge symmetry contained in $U\left( n\right) $. The general
result relating the filling factor to the $U\left( n\right) $ gauge
subgroups is summarized in the following table.

\bigskip

\begin{equation}
\begin{tabular}{|l|l|l|l|l|}
\hline Gauge Group & $U\left( 1\right) ^{n}$ & $U\left( n\right) $
& $U\left( 1\right) ^{n-j}\times U\left( n-j\right) $ & $\otimes
_{r}U\left( n_{r}\right)$\\
 &&&& \\
 \hline Filling factor &
$\sum_{a=1}^{n}\frac{1}{k_{a}}$ & $\frac{n}{k}$ & $\left(
\sum_{a=1}^{n-j}\frac{1}{k_{a}}\right) +\frac{n-j}{k}$ & $\sum_{r=1}^{n_{r}}%
\frac{n_{r}}{k_{r}}$ \\
  &&&& \\
 \hline
\end{tabular}
\end{equation}

\bigskip

From this table we learn, amongst others , that the filling fraction of a
FQH \ $n$ layers state with gauge symmetry $U\left( n-m\right) \times
U\left( m\right) $ reads in general as
\begin{equation}
\mathrm{\nu }=\frac{n-m}{k_{1}}+\frac{m}{k_{2}}.
\end{equation}
This relation tells us that the multilayer system is made of $\left(
n-m\right) $\ coincident layers with individual filling factor $1/k_{1}$ and
$m$\ coincident layers with individual filling factor $1/k_{2}$. A necessary
condition to have $U\left( n\right) $\ gauge invariance is then $k_{1}=k_{2}$%
; that is layers with same filling factors. If one insists of having layers
with different filling factors; $U\left( n\right) $\ gauge symmetry is
automatically broken down to $U\left( n\right) $ subgroups. This property
can be learnt on the way we solve the constraint eqs. Let us comment a
little bit this important point by explicit computation of the second type
of solutions.

\textbf{(2)}\textit{\ Solution II : Non abelian }$U\left( n\right) $\textit{%
\ Chern Simons model broken down to a subgroup }$G$

\qquad If one is not interested in having an exact $U\left( n\right) $
invariance, the constraint equations (3.27,\ref{contr1}) may be solved by
restricting to $\mathbf{U}$ gauge transformations in subgroups of $U\left(
n\right) $.

\textbf{(a)}\textit{\ Case} $G=U\left( 1\right) ^{n}\subset U\left( n\right)
$

\qquad In case where all layers have different filling factors, the
underlying $U\left( n\right) $ gauge group is broken down to $U\left(
1\right) ^{n}$; see figure 6. In this configuration, the $n\times n$ unitary
matrix transformation $\mathbf{U}$ have the following diagonal form,
\begin{equation*}
\mathbf{U=}\left(
\begin{array}{cccc}
\exp i\phi _{1} &  &  &  \\
& . &  &  \\
&  & . &  \\
&  &  & \exp i\phi _{n}
\end{array}
\right) ,
\end{equation*}

and so eqs(3.27,\ref{contr1}) are automatically fulfilled. However, the
current conservation condition (\ref{constr2}), although does not affect the
$\mathbf{P}$ matrix, still requires that the $\mathbf{Q}$ matrix should
satisfy,
\begin{equation}
\frac{1}{2\pi }\varepsilon ^{\mu \nu \sigma }\partial _{\mu }\mathrm{Tr}%
\left( \mathbf{Q.}\left[ \mathbf{A}_{\nu },\mathbf{A}_{\sigma }\right]
\right) =0,
\end{equation}
which is filled if and only if the matrix $\mathbf{Q}$ is proportional to
the identity. In this case, the current density $\mathbf{J}^{0}$ may be
solved, by setting $\mathbf{Q=I}$\ and $\mathbf{P=}$ $\mathrm{\kappa }$, as
\begin{equation}
\mathbf{J}^{0}=\frac{1}{2\pi }\varepsilon ^{0\nu \sigma }\left[ \mathrm{%
\kappa }\left( \partial _{\nu }\mathbf{A}_{\sigma }+i\mathrm{\kappa }^{-1}%
\left[ \mathbf{A}_{\nu },\mathbf{A}_{\sigma }\right] \right) \right] ,
\end{equation}
and the the relation $\mathcal{N}=\mathcal{\nu N}_{\phi }$ as well as its $%
SO\left( 1,2\right) $ covariance give
\begin{equation}
\frac{1}{2\pi }\varepsilon ^{\mu \nu \sigma }\left[ \mathrm{\kappa }\left(
\partial _{\nu }\mathbf{A}_{\sigma }+i\mathrm{\kappa }^{-1}\left[ \mathbf{A}%
_{\nu }\mathbf{,A}_{\sigma }\right] \right) \right] =\frac{1}{2\pi }%
\varepsilon ^{\mu \nu \sigma }\partial _{\nu }\mathcal{A}_{\sigma }.\mathrm{%
I,}
\end{equation}
which is nothing but the equation of motion one gets from the following
action
\begin{equation}
S\left[ A_{\mu },\mathcal{A}_{\mu }\right] =\frac{\mathrm{1}}{4\pi }\int
d^{3}y.\varepsilon ^{\mu \nu \sigma }\mathrm{Tr}\left( \mathrm{\kappa }%
\mathbf{A}_{\mu }D_{\nu }\mathbf{A}_{\sigma }\mathbf{-}2\mathbf{A}_{\mu }%
\mathbf{D}_{\nu }\mathcal{A}_{\sigma }\right)   \label{actionu1}
\end{equation}
where we have set $D_{\nu }=\mathrm{I.}\partial _{\nu }+\frac{2i}{3}\mathrm{%
\kappa }^{-1}\left[ \mathbf{A}_{\nu },\right] $. Under a $U\left( 1\right)
\subset U\left( n\right) $ gauge transformation
\begin{equation}
\mathbf{A}_{\sigma }^{\prime }=\mathbf{A}_{\sigma }+\partial _{\sigma
}\lambda .\mathrm{I;\quad \quad }\left[ \mathbf{A}_{\nu },\partial _{\sigma
}\lambda .\mathrm{I}\right] =0
\end{equation}
the terms appearing in the above action, namely $D_{v}\mathbf{A}_{\sigma }$\
and $\mathbf{A}_{\mu }D_{\nu }\mathbf{A}_{\sigma }$, transform as
\begin{eqnarray}
D_{v}^{\prime }\mathbf{A}_{\sigma }^{\prime } &=&D_{v}\mathbf{A}_{\sigma }+%
\mathrm{I.}\partial _{\nu }\partial _{\sigma }\lambda ,  \notag \\
\mathbf{A}_{\mu }^{\prime }D_{\nu }^{\prime }\mathbf{A}_{\sigma }^{\prime }
&=&\mathbf{A}_{\mu }D_{\nu }\mathbf{A}_{\sigma }+\mathrm{I.}\partial _{\mu
}\partial _{\nu }\partial _{\sigma }\lambda ,
\end{eqnarray}
and so invariance of the functional eq(\ref{actionu1}) is ensured by the
completely antisymmetric factor $\varepsilon ^{\mu \nu \sigma }$. The
filling fraction formula
\begin{equation}
\mathrm{\nu }=\mathrm{Tr}\left( \mathrm{\kappa }^{-1}\right)   \label{ff2}
\end{equation}
for the $U\left( 1\right) ^{n}$ model with interaction is similar to the one
we have got earlier in the case of the multilayer system without
interactions. But one should note that while $\mathrm{\kappa }$\ is diagonal
in the last case; it is however not for the first one allowing extra
contributions\footnote{%
Here we represent the link between the coupling matrix $\mathrm{\kappa }$\
and gauge symmetry of the layers in the limit $\left| d_{a}-d_{b}\right|
\rightarrow 0$, $\forall a,b=1,...,n$. This diagram should not be confused
with the geometric one of figure 1 representing the base manifold.}; see
figure 6.

\bigskip

\begin{figure}[tbh]
\begin{equation*}
\epsfxsize=6cm\epsffile{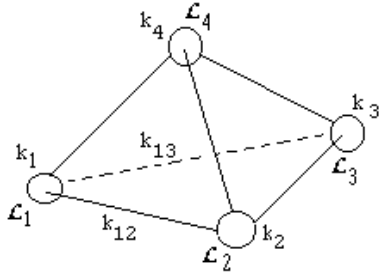}
\end{equation*}
\caption{\textit{\ This figure represents a system of }$n=4$\textit{\ layers
with mutual interactions. The general graph on gets is a simplex of four
vertices and six edges. The coupling matrix has in general ten free integers}
$k_{ab}=k_{ba}$ \textit{\ and the filling fraction is given by eq(3.47).
Subsystems of this configuration are obtained by making choices on the
integers} $k_{ab}$ \textit{\ such as for instance identifying the quantum
numbers carried by two vertices and their link.}}
\label{1}
\end{figure}

\textbf{(b)}\textit{\ }\ \textit{General Case} $G\subset U\left( n\right) $

\qquad To see how the solution of the constraint eqs (3.27,\ref{contr1},\ref
{constr2}) for generic subgroups of $U\left( n\right) $, let us consider the
example where $U\left( n\right) $ is broken down to $U\left( 2\right) \times
U\left( 1\right) ^{n-2}$. In this case, the $n\times n$ unitary matrix
transformation $\mathbf{U}$ have the following diagonal form
\begin{equation*}
\mathbf{U=}\left(
\begin{array}{cccccc}
u_{11} & u_{12} & 0 &  &  & 0 \\
u_{21} & u_{22} & 0 &  &  & 0 \\
0 & 0 & \exp i\phi _{1} &  &  &  \\
&  &  & . &  &  \\
&  &  &  & . &  \\
0 & 0 &  &  &  & \exp i\phi _{n-2}
\end{array}
\right) ,
\end{equation*}
where $u_{ij}$, $\left| u_{11}u_{22}-u_{12}u_{21}\right| =1$, are elements
of $U\left( 2\right) $ and $\phi _{a}$ phase of $U\left( 1\right) ^{n-2}$.
While the matrix $\mathbf{Q}$ is usually proportional to the identity, the
matrix $\mathbf{P=}$ $\mathrm{\kappa }$ should the form

\begin{equation*}
\mathrm{\kappa =}\left(
\begin{array}{ccc}
k & 0 & k_{1a} \\
0 & k & k_{2a} \\
k_{1a} & k_{2a} & \mathrm{\kappa }^{\prime }
\end{array}
\right)
\end{equation*}
where $\mathrm{\kappa }^{\prime }$ is a $\left( n-2\right) \times \left(
n-2\right) $\ hermitian matrix. By repeating this analysis, one gets all
possible configurations. For the special case where $k_{1a}=k_{2a}=0$, one
has a reducible diagram.

\begin{itemize}
\item  \textsl{Examples}
\end{itemize}

\qquad To illustrate the prediction of broken $U\left( n\right) $ gauge
models, let us consider two examples; the first one concerns two layers FQH
states and the second one involves three layers.

\textbf{(a) }{\large Two layers States}

In the example of two coupled layers $\mathcal{L}_{1}$ and $\mathcal{L}_{2}$
with the hermitian coupling matrix $\mathrm{\kappa }$, reads as
\begin{equation}
\mathrm{\kappa =}\left(
\begin{array}{cc}
k_{1} & l \\
l & k_{2}
\end{array}
\right)
\end{equation}
where $k_{1}$, $k_{2}$ are odd integers and, a priori, $l$ is any positive
integer; see figure 7

\bigskip

\begin{figure}[h]
\begin{equation*}
\vspace{2cm}\epsfxsize=7cm\epsffile{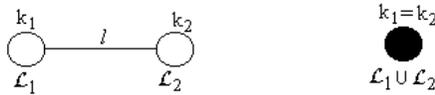}
\end{equation*}
\caption{\textit{This is the quiver diagram of two layers FQH states. For
different values of } $k_{1}$ \textit{and } $k_{2}$ and the field model has
a $U(1)^{2}$ abelian symmetry \textit{\ the total filling factor of the
system is given by eq(3.48). Non abelian } $U(2)$ \textit{\ model is
obtained for} $k_{1}=k_{2}=k$ \textit{and } $k_{12}=0$. }
\label{figure 7}
\end{figure}

In this case filling factor $\mathrm{\nu }$, that follows from eq(\ref{ff2}%
), is given by the following three integer series,
\begin{equation}
\mathrm{\nu }=\frac{k_{1}+k_{2}}{k_{1}k_{2}-l^{2}}
\end{equation}
This is a general relation containing as special cases known results on FQH
states with two levels. Setting $l=0$, one gets the familiar relation $%
\mathrm{\nu }=1/k_{1}+1/k_{2}$ of uncoupled $U\left( 1\right) ^{2}$ FQH
states which gets an enhanced $U\left( 2\right) $ symmetry for $%
k_{1}=k_{2}=k $. One may also recover the Jain series $2/\left( 4p\pm
1\right) $ at level two by taking
\begin{equation}
k_{1}=k_{2}=k;\quad l^{2}=nk;\quad k-n=4p\pm 1
\end{equation}
As an explicit example, one sees that the solutions for the state $\mathrm{%
\nu }=2/5$ reproduce all known results. Expressing $l$ in terms of $k_{1}$
and $k_{2}$, one gets a constraint relation on the values of $l$ integers
one should have for $\mathrm{\nu }=2/5$,
\begin{equation}
l=\frac{1}{2}\sqrt{\left( 2k_{1}-5\right) \left( 2k_{2}-5\right) -25};\quad
l\in Z^{+}.
\end{equation}
The solutions are as follows: (i) $k_{1}=k_{2}=k$ and $l=\sqrt{k\left(
k-5\right) }$ which is fulfilled for $l=0$ and $k=5$ ; but also $l=6$ and $%
k=9$. The first solution corresponds to the multilayer system$\left(
550\right) $ in terms of the conventional classification made in \cite{22};
while the second corresponds to $\left( 332\right) $. (ii) There is also an
other solution for $k_{1}=3$, $k_{2}=15$ and $l=0$ in a complete agreement
with Haldane result, to be discussed later on. The same is also valid for $%
\mathrm{\nu }=2/3$\ where the integrality condition on $l$ reads as
\begin{equation}
l=\frac{1}{2}\sqrt{\left( 2k_{1}-3\right) \left( 2k_{2}-3\right) -9};\quad
l\in Z^{+}.
\end{equation}
A general result can be written down for all terms of the Jain sequence; see
eq(1.6).

\textbf{(b)}\textit{\ }{\large Three layers States}

For the case of three layers, the matrix $\mathrm{\kappa }$\ has in general
three odd $k_{a}$ integers associated with self interactions and three
other\ integers $l$, $m$, $t$ associated with the three kinds of couplings;
see figure 8.
\begin{equation}
\mathrm{\kappa =}\left(
\begin{array}{ccc}
k_{1} & l & m \\
l & k_{2} & t \\
m & t & k_{3}
\end{array}
\right) .
\end{equation}
This is a real symmetric matrix whose diagonal elements $k_{a}$ are positive
odd integers giving the filling factors of the layers in lowest Landau level
configuration. The off diagonal entries encode layers interactions.

\begin{figure}[h]
\vspace{1.5cm}
\begin{equation*}
\epsfxsize=4cm\epsffile{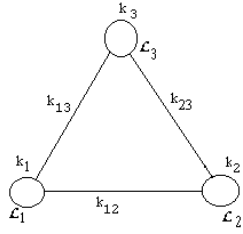}
\end{equation*}
\caption{\textit{Here we give the diagram for FQH states involving three
layers in the first Landau level. The filling fraction is given by eq(3.53).
For }$k_{a}\neq k_{b}\neq k_{c}$ and $k_{ab}=k_{ac}=k_{bc}=0$\textit{\ \ we
have an abelian }$U(1)^{3}$\textit{\ symmetry without layers interactions.
In the other cases there are interactions and in general the symmetry may be
non abelian. In the case } $k_{a}=k_{b}=k_{c}$ and $k_{ab}=k_{ac}=k_{bc}=0$,%
\textit{\ we have the largest symmetric }$U(3)$ \textit{non abelian gauge
model.}}
\label{figure 8}
\end{figure}
This matrix depends on $\left( 3+3\right) =6$ integers allowing to classify
the three kinds of effective field models one has in this case; see figure 9.

$\left( \alpha \right) $ \textit{Model with }$U\left( 1\right) ^{3}$\textit{%
\ symmetry.}

Here, the $k_{a}$ integers are different and $l$, $m$ and $t$ are generally
non zero. The filling factor one gets depends then on $\left( 3+3\right) $
integers and reads in general as,
\begin{equation}
\mathrm{\nu }=\frac{\left( k_{2}k_{3}+k_{1}k_{3}+k_{1}k_{2}\right) -\left(
l^{2}+m^{2}+t^{2}\right) }{k_{1}k_{2}k_{3}+2lmt-\left(
k_{1}t^{2}+l^{2}k_{3}+m^{2}k_{2}\right) }
\end{equation}
Special cases may be also considered here; like for instance retaining only
couplings dealing with closer neighboring layers by setting $m=0$ or again
by setting $l=t$ and $m=0$.

\bigskip

\begin{figure}[tbh]
\begin{equation*}
\epsfxsize=12cm\epsffile{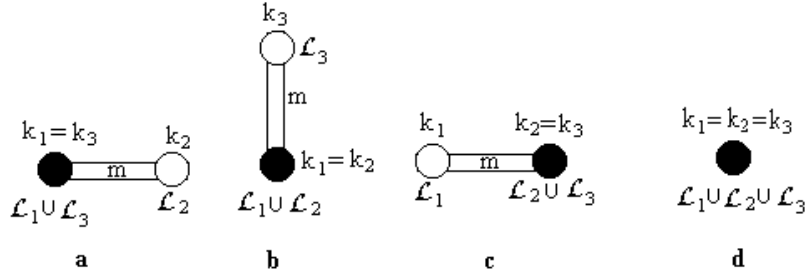}
\end{equation*}
\caption{\textit{Figures 9a, 9b and 9c are obtained from figure 8 by
coinciding two nodes giving then a enhanced }$U\left( 2\right) $\textit{\
symmetry. For the case }$k_{a}=k_{b}\neq k_{c}$, $U\left( 2\right) $\textit{%
\ symmetry requires }$k_{ab}=0$\textit{\ but }$k_{ac}=k_{bc}$. In figure 9d,
the three nodes are coincident and the symmetry is $U\left( 3\right) $.}
\label{1}
\end{figure}

$\left( \beta \right) $ \textit{Models with }$U\left( 1\right) \times
U\left( 2\right) $\textit{\ symmetry.}

Here, two of the three $k_{a}$ integers are equal and one of the $l$, $m$
and $t$ integers is zero and the remaining two others are equal. For
instance, we can have $k_{1}=k_{2}=k$ and $l=0$ together with $m=t$. In this
configuration, which is associated with the diagram of figure 9b, the
filling factor one gets depends then on $3$ integers $k$, $k_{3}$ and $m$
and reads in general as,
\begin{equation}
\mathrm{\nu }=\frac{k^{2}-2m^{2}+2kk_{3}}{k^{2}k_{3}-2km^{2}}=\mathrm{\nu }%
_{U\left( 2\right) }+\mathrm{\nu }_{U\left( 1\right) }
\end{equation}
This relation may also be decomposed as $\mathrm{\nu }=\mathrm{\nu }%
_{U\left( 2\right) }+\mathrm{\nu }_{U\left( 1\right) }$ with,
\begin{eqnarray}
\mathrm{\nu }_{U\left( 2\right) } &=&\frac{2k_{3}}{kk_{3}-2m^{2}}  \notag \\
\mathrm{\nu }_{U\left( 1\right) } &=&\frac{k^{2}-2m^{2}}{k^{2}k_{3}-2km^{2}}
\end{eqnarray}
In absence of layer interactions; that is for m=0; the above eqs reduce to
the standard situation where one of the layers is far away and the other two
are coincident. The last configuration one can also have corresponds to the
special case where $k_{1}=k_{2}=k_{3}=k$ and $l=m=t=0$; see figure 9d. The
symmetry of this FQH state is the full $U\left( 3\right) $ gauge invariance.

Having given the main idea on the commutative effective field of multilayer
FQH states with rational values, we pass now to make some comments regarding
the wave function approach, a matter of having a global view on the
connection between gauge symmetry, the various generalizations of the
Laughlin wave functions and their filling factors.

\subsection{Wave functions}

Here we give the generalizations of the Laughlin wave function for uncoupled
and coupled layers using the correspondence layer $\mathcal{L}%
\longleftrightarrow D2$ brane and the classification in terms of subgroups
of $U\left( n\right) $ gauge symmetry.

\subsubsection{Uncoupled Layers: $U\left( 1\right) ^{n}$ model I}

\qquad For the case where the $n$ layers are enough distant from each
others, $\left| d_{a+1}-d_{a}\right| \equiv \epsilon _{a}>>0$, that is in
absence of layer interactions, the total wave function $\Psi _{f}$
describing the ground state of the multilayer system is given by the
antisymmetric product of individual Laughlin trial waves $\Psi _{La}=\Psi
_{L}\left( z_{\alpha _{a}}\right) $
\begin{equation}
\Psi _{f}=\prod_{a=1}^{n}\Psi _{La},
\end{equation}
where the $\Psi _{La}$\ factors are given by,
\begin{equation}
\Psi _{La}=\prod\limits_{\alpha _{a}<\beta _{a}=1}^{N_{a}}\left( z_{\alpha
_{a}}-z_{\beta _{a}}\right) ^{k_{a}}\exp \left( -\frac{\mathcal{B}_{ex}}{4}%
\sum_{\gamma _{a}=1}^{N_{a}}\left| z_{\gamma _{a}}\right| ^{2}\right) ,
\label{fregenlaugh}
\end{equation}
and where $z_{\alpha _{a}}$, $\alpha _{a}=1,...,N_{a}$ are the complex
coordinates of electrons on the $a-th$ \ Laughlin layer $\mathcal{L}_{a}$.
Note that antisymmetry of the wave function $\Psi _{f}$ under the change of
any pair of electrons requires that all $k_{a}$'s to be odd integer. In this
case the filling factor is $\mathrm{\nu }=1/k_{1}+...+1/k_{n}$. The quiver
diagram describing this abelian $U\left( 1\right) ^{n}$ model is given by
figure 5.

\subsubsection{Coupled Layers:}

According to the $\left| d_{b}-d_{a}\right| $ distances between layers $%
\mathcal{L}_{a}$ and $\mathcal{L}_{b}$,$\ $one can write down different
kinds of generalized Laughlin wave function for multilayer system. Here we
give the two special ones associated with the field models eqs(3.43) and
(3.37). The same reasoning applies to the other configurations. From the
analysis of subsection 3.1, we distinguish, amongst others, the following
particular models

(1) $U\left( 1\right) ^{n}${\large \ model II: Layers with different filling
factors}.

\qquad For small distances between the layers; that is for $\mathrm{\epsilon
=}\left| d_{b}-d_{a}\right| <<1$, layer couplings are no longer negligible
and the generalized Laughlin wave function $\Psi _{coupled}$ describing the
ground state of the multilayer system takes a more complicated form. Since
the filling factors are not equal, $k_{1}\neq k_{2}\neq ...\neq k_{n}$, the $%
U\left( n\right) $ gauge symmetry, which exist in case where all $k$'s are
equal, is now broken down to $U\left( 1\right) ^{n}$; see figures 6 for
illustration.

This model involves, besides the diagonal terms, monomials type $\left(
z_{\alpha _{a}}-z_{\beta _{b}}\right) ^{\mathrm{k}_{ab}}$ with $a\neq b$ and
$\mathrm{k}_{ab}$\ some integers. The general form of $\Psi _{coupled}$
extending the Laughlin wave function read then as,
\begin{equation}
\Psi _{coupled}=\prod\limits_{a\leq b=1}^{n}\prod\limits_{\alpha _{a}<\beta
_{a}=1}^{N_{a}}\left( z_{\alpha _{a}}-z_{\beta _{b}}\right) ^{\mathrm{k}%
_{ab}}\exp \left( -\frac{B}{4}\sum_{a=1}^{n}\sum_{\gamma
_{a}=1}^{N_{a}}\left| z_{\gamma _{a}}\right| ^{2}\right)
\end{equation}

As the layers are closed to each others, electrons may travel from a layer $%
\mathcal{L}_{a}$ to an other $\mathcal{L}_{b}$ and so antisymmetry, of the
wave function under the change of any pair $\left( z_{\alpha _{a}},z_{\beta
_{b}}\right) $ of electron coordinates, is no longer a constraint since
layers are different. Therefore, the matrix entries $\mathrm{k}_{ab}$ are
not required to be odd integers and the total filling factor is given by eq(
\ref{ff2}). In $n\times n$ matrix notation, the coordinates variables $%
z_{\alpha _{a}}$ on the $a-th$ layer $\mathcal{L}_{a}$ may be thought of as
the $\left( a,a\right) $ diagonal element of $n\times n$ matrix $\mathbf{Z}%
_{\alpha },$
\begin{equation*}
z_{\alpha _{a}}=<a|\mathbf{Z}_{\alpha }\mathbf{|}a\mathbf{>}
\end{equation*}
while the coordinates $z_{\alpha _{ab}}$ of particles traveling from $%
\mathcal{L}_{a}$ to $\mathcal{L}_{b}$ are just,
\begin{equation*}
z_{\alpha _{ab}}=<b|\mathbf{Z}_{\alpha }\mathbf{|}a\mathbf{>}
\end{equation*}
One can use this property to study the cases where the generalized Laughlin
wave functions have non abelian symmetries containing the underlying $%
U\left( n\right) $ gauge symmetry of the coincident $D2$ branes. In what
follows, we shall give details for the case of FQH states with full $U\left(
n\right) $ symmetry.

(2) {\large Non Abelian }$U\left( n\right) ${\large \ model}

This happens in the case where all layers $\mathcal{L}_{a}$, viewed as $D2$
branes, have the same filling factor $1/k$, $k$\ odd integer, the total
number $\mathcal{N}$ of electrons moving on $\left\{ \cup _{a}\mathcal{L}%
_{a}\right\} $ may, at a given time, be partitioned into $N_{a}$ particles
moving on $\mathcal{L}_{a}$, \textit{plus} $N_{ab}$ particles leaving $%
\mathcal{L}_{a}$ for $\mathcal{L}_{b}$\ and $N_{ba}$ electrons coming from $%
\mathcal{L}_{b}$ and landing on $\mathcal{L}_{a}$. These particles form
together a $n\times n$ matrix $\mathbf{N}$ with positive integer entries,
\begin{equation}
\mathbf{N}=\left( N_{ab}\right)
\end{equation}
Note that the total number $\mathcal{N}$ of electrons is given by the trace
of this matrix, namely $\mathcal{N}=\mathrm{Tr}\left( \mathbf{N}\right)
=N_{1}+...+N_{n}$; see also eqs (3.18-3.21). Moreover as the layers have the
same filling factor $1/k$; it follows from $U\left( n\right) $\ invariance
that, at any time, $N_{1}=...=N_{n}=N$; and so $\mathcal{N}=nN$.

Since on a given layer $\mathcal{L}_{a}$ each electron has a coordinate $%
z_{\alpha _{a}}$, one can parameterize collectively these particles in a
more convenient way by using $n\times n$ matrices. To get the $U\left(
n\right) $ invariant Laughlin wave function, we start from eq(\ref
{fregenlaugh}) with $k_{1}=...=k_{n}=k$. Then instead of diagonal matrices,
we consider a system of $N$ matrices matrices $\left\{ \mathbf{Z}_{\alpha
}\right\} $ with general entries
\begin{equation}
\mathbf{Z}_{\alpha }=\left( z_{\alpha _{ab}}\right) ;\qquad a,b=1,...,n
\end{equation}
where in addition to $z_{\alpha _{aa}}$, the complex coordinates of
electrons moving on $\mathcal{L}_{a};$ we have also $z_{\alpha _{ab}}$\ and $%
z_{\alpha _{ba}}$ variables describing respectively the coordinates of the
electrons leaving $\mathcal{L}_{a}$ for $\mathcal{L}_{b}$\ and the particles
coming from $\mathcal{L}_{b}$ to $\mathcal{L}_{a}$. The generalized wave
function, taking into account multilayers interactions and invariance under $%
U\left( n\right) $ symmetry, reads as,

\begin{equation}
\Psi _{U\left( n\right) }=\prod\limits_{\alpha <\beta =1}^{N}\left[ \det
\left( \mathbf{Z}_{\alpha }-\mathbf{Z}_{\beta }\right) \right] ^{\mathrm{k}%
}\exp \left( -\frac{\mathcal{B}_{ex}}{4}\sum_{\gamma =1}^{N}\mathrm{Tr}%
\left( \mathbf{Z}_{\gamma }\mathbf{Z}_{\gamma }^{\dagger }\right) \right) .
\label{uninv}
\end{equation}
Acting by the following $U\left( n\right) $ gauge transformations,
\begin{equation}
\mathbf{Z}_{\alpha }^{\prime }=\mathbf{U}^{-1}\mathbf{Z}_{\alpha }\mathbf{U},
\end{equation}
invariance of eq(\ref{uninv}) follows naturally from the global action of
the transformations as well as the cyclic property of $\mathrm{det}$ and $%
\mathrm{trace}$. The filling factor is, in this special case, given by $%
\mathrm{\nu }=n/k.$

\section{NC effective field and matrix models for Multilayer States}

We first review briefly the Susskind idea for the case of a single layer
with filling factor $1/k$; then we study its extension to the case of the
multilayer system in lowest energy configuration. Hierarchies in multilayer
states based on non commutative geometry which will be discussed in section
7.

\subsection{General on Susskind NC theory}

\qquad To begin consider an electric charged particle, say an electron,
moving in the real plane in presence of an external constant magnetic field $%
\mathcal{B}_{ex}$. Classically this particle is parameterized by its
position $x^{i}(t);\ i=1,2$ and velocity $v^{i}=\partial _{t}{\ x^{i}}$. For
a $\mathcal{B}_{ex}$ field strong enough, the dynamics of the particle is
mainly governed by the coupling
\begin{equation}
S[x]=\frac{e\mathcal{B}_{ex}}{2}\int dt{\ \varepsilon _{ij}v^{i}x^{j}},
\end{equation}
which is just the angular momentum of the system, induces at the quantum
level a non commutative structure on the real plane; i.e
\begin{equation}
\lbrack x^{i},x^{j}]\propto \epsilon ^{ij}/\mathcal{B}_{ex}.
\end{equation}
If we suppose that the total number of electrons $\mathcal{N}_{e}$ is a
constant of motion; see footnote 4, thing that we shall use everywhere here,
one can write it, in the large $\mathcal{N}_{e}$ limit, as,
\begin{equation}
\mathcal{N}_{e}=\int d^{2}x\mathrm{\rho }=\int d^{2}y\text{ }\mathrm{\rho }%
_{0},
\end{equation}
where $\mathrm{\rho }$ and $\mathrm{\rho }_{0}$
\begin{equation}
\mathrm{\rho }_{0}=\left| \frac{\partial ^{2}\left( x^{1},x^{2}\right) }{%
\partial ^{2}\left( y^{1},y^{2}\right) }\right| \mathrm{\rho }
\end{equation}
\ are the densities of particles in the $\left\{ x^{i}\right\} $ and $%
\left\{ y^{i}\right\} $\ frames respectively.

In the case of $\mathcal{N}_{e}$ classical particles, without mutual
interactions, parameterized by the coordinates $x_{a}^{i}(t)$ and velocities
$v_{a}^{i}$, the dynamics is dominated by the $B-x(t)$ couplings extending
eq(4.1) as
\begin{equation}
\frac{e\mathcal{B}_{ex}}{2}\int dt\varepsilon _{ij}\ \Sigma
_{a=1}^{N}v_{a}^{i}x_{a}^{j},  \label{genaction}
\end{equation}
and describing a typical strongly correlated system of electrons showing a
quantum Hall effect with filling fraction
\begin{equation}
\mathrm{\nu }=\frac{\mathcal{N}_{e}}{\mathcal{N}_{\phi }};\quad \mathrm{\nu }%
.\mathcal{N}_{\phi }=\mathcal{N}_{e}.  \label{laugh10}
\end{equation}
Expressing the number $\mathcal{N}_{\phi }$ of quantum flux, by using the
external magnetic field, as
\begin{equation}
\mathcal{N}_{\phi }=\frac{1}{2\pi }\int_{\mathcal{L}}d^{2}y\text{ }\mathcal{B%
}_{ex}=\frac{1}{2\pi }\oint_{\partial \mathcal{L}}\mathcal{A}^{ex}dl
\end{equation}
where $\mathcal{B}_{ex}=\varepsilon ^{3\mu \nu }\partial _{\mu }\mathcal{A}%
_{\nu }^{ex}$ and using eq(\ref{laugh10}), one gets relation the density $%
\mathrm{\rho }_{0}$ of electrons and quantum fluxes, namely $\mathrm{\rho }%
_{0}=\mathrm{\nu }\frac{\mathcal{B}_{ex}}{2\pi }$.
\begin{figure}[h]
\vspace{1.5cm}
\begin{equation*}
\epsfxsize=3cm \epsffile{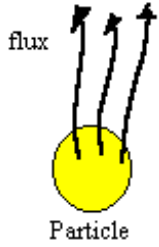}
\end{equation*}
\caption{In the Laughlin picture, each electron is coupled with k quantum
fluxes as shown on the figure with k=3 . The relation $\protect\nu {\mathcal{%
N}_{\protect\phi }}={\mathcal{N}_{e}}$ reflects the identity between the
densities of electrons and quantum fluxes. As such the conservation of the
total number of electrons may be thought of as the charge of the Noether
conserved current generating the diffeomorphism symmetry of the y plane or
equivalently the gauge symmetry of the Susskind NC effective field model.}
\label{figure 9}
\end{figure}

Quantum mechanically, there are different field theoretical methods to
approach the quantum states of this system, either by using techniques of
non relativistic quantum mechanics \cite{laughwf}, methods of conformal
field theory especially for the study of droplets and edge excitations \cite
{23',23'''} or again by using the CS effective field model \cite{23''}
describing the limit $\mathcal{N}\rightarrow \infty $ of electrons where now
the Hall system is viewed as a liquid of particles. In this case, the Chern
Simons theory in the $(2+1)$ dimensional space modeling the FQH Laughlin
state of filling fraction $\nu =\frac{1}{k}$ is obtained by interpreting the
$\mathrm{\rho }_{0}$ density as
\begin{equation}
\mathrm{\rho }_{0}=\mathcal{J}^{0}\left( y\right) ,
\end{equation}
the time component of a $\left( 1+2\right) $ conserved Noether current $%
\mathcal{J}^{\mu }\left( y\right) $, which is realized, in terms of $\left(
1+2\right) $ Chern Simons gauge field $A_{\sigma }$,
\begin{equation}
\mathcal{J}^{\mu }\left( y\right) =\frac{\mathrm{1}}{2\pi }\varepsilon ^{\mu
\nu \sigma }\partial _{\nu }A_{\sigma },
\end{equation}
In terms of this gauge field representation, the number of electrons reads
as,
\begin{equation}
\mathcal{N}=\int_{\mathcal{L}}B=\frac{1}{2\pi }\oint_{\partial \mathcal{L}%
}Adl,
\end{equation}
where now $B$ is the gauge invariant two form field associated with the CS
gauge potential $A$. In terms of this $A$ field, the action describing
electrons in presence of the external potential $\mathcal{A}^{ex}$ reads
then as,
\begin{equation}
S[A]=\frac{\mathrm{k}}{4\pi }\int d^{3}y\text{ }\varepsilon ^{\mu \nu \rho
}\partial _{\mu }A_{\nu }A_{\rho }-\frac{1}{2\pi }\int d^{3}yJ^{\mu }%
\mathcal{A}_{\mu },  \label{actionl}
\end{equation}
Variation of this action functional with respect to $A_{\rho }$, one gets
\begin{equation}
\mathrm{k}\epsilon ^{\mu \nu \rho }\partial _{\mu }A_{\nu }=\epsilon ^{\mu
\nu \rho }\partial _{\mu }\mathcal{A}_{\nu },
\end{equation}
which, after integration of the time component \ of this eq over the layer,
one discovers the relation (\ref{laugh10}) between the numbers $\mathcal{N}$
and $\mathcal{N}_{\phi }$.

The link between this field action and FQH fluids dynamics has been studied
in details and most of the results in this direction has been established
several years ago \cite{23'',27,28,281,29,291,292}. However an interesting
observation has been made few years ago by Susskind \cite{1} and further
considered in \cite{31,32,322,323,33,331,34,35,36,361}. The novelty brought
by the study made in \cite{1} is that: (1) Because of the $\mathcal{B}_{ex}$%
-field, level $\mathrm{k}$ NC Chern-Simons $U(1)$ gauge theory may provide a
description of the Laughlin theory at filling fraction $\nu _{L}=\frac{1}{%
\mathrm{k}}$. In this vision, Eq(\ref{actionl}) appears then just as the
leading term of a more general theory which reads as:
\begin{eqnarray}
S[A] &=&\frac{1}{4\pi \mathrm{\nu }}\int d^{3}y\epsilon ^{\mu \nu \rho
}[\partial _{\mu }A_{\nu }\ast A_{\rho }+\frac{2i}{3}A_{\mu }\ast A_{\nu
}\ast A_{\rho }]  \label{ncactio} \\
&&-\frac{1}{2\pi }\int d^{3}yJ^{\mu }\mathcal{A}_{\mu },  \notag
\end{eqnarray}
where $\ast $ stands for the usual star operation of non commutative field
theory \cite{37}. (2) The above NC Chern Simons $U(1)$ action is in fact the
$\mathcal{N}\rightarrow \infty $ of the following finite dimension matrix
model action:
\begin{eqnarray}
S\left[ Z,\Psi ,A\right] &=&{\frac{\mathrm{k}}{4\theta }}\int dtTr\left( i{%
\bar{Z}}DZ-\omega Z{\bar{Z}}\right) +\ hc  \label{matrixsp} \\
&&\frac{i}{2}\int dt\Psi ^{\dagger }D\Psi +\frac{\mathrm{k}}{2}\int dtTr%
\mathrm{A}+\ hc.  \notag
\end{eqnarray}
In this relation $Z=X^{1}+iX^{2}$, $\Psi $ is the Polychronakos field which
play the role of a regulator; but has a remarkable interpretation in brane
language, and $\mathrm{A}$ is a Lagrange field multiplier carrying the
constraint of the system. The potential $\omega Z{\bar{Z}}$ is introduced in
order to keep the electrons near the origin; but in our present study one
may ignore it. In the large $\mathcal{N}$ limit the variable fields $x\left(
t\right) $ are mapped to $\left( 1+2\right) $\ ones $X\left( y\right) $;
which upon substituting the following Susskind mapping,
\begin{eqnarray}
X^{i} &=&y^{i}+\theta \varepsilon ^{ij}A_{j},  \label{sussmap} \\
\lbrack y^{i},y^{j}] &=&i\theta \varepsilon ^{ij}.
\end{eqnarray}
into eq(\ref{genaction}), one discovers the leading order of the $\theta $\
expansion of eq (\ref{ncactio}).

The analysis we have given here above concerns the Laughlin state; the
ground state configuration of FQH systems. In what follows we want to
generalize this analysis for the case of FQH states with rational filling
factor.

\subsection{ Extension for Multilayer States}

\qquad One of the key points in the NC Chern Simons effective field approach
of the Laughlin ground state\ is the Susskind map, which reads, for the case
of single layer system, as in eq(\ref{sussmap}), where $\theta $ is the
usual NC parameter of the Moyal plane. In the multilayer case where we have $%
n$ parallel $D2$ branes, the above map should be thought of as,
\begin{equation}
x_{1}^{i}=y_{1}^{i}+\theta _{1}\varepsilon ^{ij}\mathrm{A}_{j}^{1}\left(
y\right) ,
\end{equation}
the map associated with the layer $\mathcal{L}_{1}$ parameterized by the
real coordinates.
\begin{equation}
\mathcal{L}_{1}:\left\{ y_{1}^{1},y_{1}^{2};\quad y^{3}=d_{1}\right\}
\end{equation}
In complex notations $z_{1}=x_{1}^{1}+ix_{1}^{2}$ and $w_{1}=\left(
y_{1}^{1}+iy_{1}^{2}\right) $, this splitting reduces to a more simple form
\begin{equation}
z_{1}=w_{1}+\theta _{1}\mathrm{A}_{\overline{w}}^{1}
\end{equation}
where $\mathrm{A}_{\overline{w}}^{1}\sim \left( \mathrm{A}_{1}^{1}-i\mathrm{A%
}_{2}^{1}\right) $. In the case of a FQH system $\left\{ \cup _{a}\mathcal{L}%
_{a}\right\} $ with $n$ layers parameterized by $\left\{ w_{a};\quad
y^{3}=d_{a}\right\} $; the general form extending the Susskind map, one may
write down, is
\begin{eqnarray}
z_{a} &=&w_{a}+\sum_{b=1}^{n}\theta _{ab}\mathrm{A}_{\overline{w}}^{ba}
\label{sussgen} \\
x^{3} &=&y^{3}=d_{a}  \notag
\end{eqnarray}
where now the extra index $a$ refers to the $n$ parallel layers of the
system and where the CS gauge fields $\mathrm{A}_{\overline{w}}^{ba}$ carry
a dependence on the layer spacing $d_{a}$ moduli; that is,
\begin{equation}
\mathrm{A}_{\overline{w}}^{ba}=\mathrm{A}_{\overline{w}}^{ba}\left( w,%
\overline{w},y^{3}\right) .
\end{equation}
Since layers are parallel, one may here also distinguish different
configurations in one to one correspondence with the $U\left( n\right) $
symmetry subgroups. These configurations correspond also to the various
possibilities one has regarding the choices of the NC $\theta _{ab}$
parameters. A natural way to link $\theta _{ab}$\ and $k_{ab}$ integers \ is
as,
\begin{equation}
\left[ w_{a},\overline{w}_{b}\right] =-2\theta _{ab}=2i\frac{k_{ab}}{B}.
\end{equation}
We have already studied the general case where $k_{ab}$ is an invertible
matrix; in particular the link with subgroups of\ $U\left( n\right) $; here
we shall focus our attention on the more symmetric case where the $z_{a}$
and $w_{a}$ variables are realized as,
\begin{equation}
z_{a}=z.\mathbf{\pi }_{a};\quad w_{a}=w.\mathbf{\pi }_{a},
\end{equation}
with $\mathbf{\pi }_{a}$\ are the projectors on the $a-th$ states introduced
in section 3 eq(3.9). This representation satisfies
\begin{equation}
\left[ w.\mathbf{\pi }_{a},\overline{w}.\mathbf{\pi }_{a}\right] =-2\theta .%
\mathbf{\pi }_{a}=2i\frac{k}{B}\mathbf{\pi }_{a}
\end{equation}
where $k$ is an odd integer.

Moreover as the layers are identified with $D2$ branes, one may think about
the generalized Susskind map fluctuations $\mathrm{A}_{\overline{w}%
}^{ba}\left( w,\overline{w},y^{3}\right) $ of eq(\ref{sussgen}) as
describing quantum fluctuations of fundamental strings stretching between $%
\mathcal{L}_{a}$ and $\mathcal{L}_{b}$ layers. As such, one can represent
collectively the $n^{2}$ functions $\mathrm{A}_{\overline{w}}^{ba}\left( w,%
\overline{w},y^{3}\right) $ in terms of $n\times n$ hermitian matrix as,
\begin{equation}
\mathrm{A}_{\overline{w}}\left( w,\overline{w},d_{b}\right) \sim \mathbf{\pi
}_{a}\sum_{a}\mathrm{b}_{\overline{w}}^{a}\left( w,\overline{w}\right) +%
\frac{1}{1+\frac{d_{b}-d_{a}}{d_{a}}}\text{ }\mathbf{u}_{ba}\text{ }\mathrm{b%
}_{\overline{w}}^{ba}\left( w,\overline{w};d_{a}\right)  \label{ddepend}
\end{equation}
where $\mathbf{u}_{ab}=|a><b|$. To have an idea on the way we have derived
this approximation for the layers gauge fields; set $y^{3}=\tau $, the
fields $\mathrm{b}_{\overline{w}}^{a}\left( w,\overline{w},d_{a}\right)
=F\left( \tau \right) $,
\begin{equation}
F\left( \tau \right) =\int \frac{dp}{2\pi }\widetilde{F}\left( p\right) \exp
ip\tau ;
\end{equation}
and $d_{a}=\tau _{a}$, ordered as
\begin{equation}
0=\tau _{1}\leq \tau _{1}\leq ...\leq \tau _{n}.
\end{equation}
Then use the limit $\tau _{ab}$ small and the Heisenberg incertainty
relation, $p\tau \sim 1$, to expand $F\left( \tau \right) $ as\
\begin{equation}
F\left( \tau _{n}\right) \sim \left( 1-\frac{\tau _{n}-\tau _{n-1}}{\tau
_{n-1}}\right) F\left( \tau _{n-1}\right) \sim \left( 1-\varepsilon \right)
F\left( \tau _{n-1}\right) ,  \label{approx}
\end{equation}
where $\varepsilon $\ is a small fluctuation. Note that this approximation,
which is valid for $\tau _{n}-\tau _{n-1}<<<1$, may be prolonged to the
region \ $\tau _{n}-\tau _{n-1}>>>1$\ just by thinking about $\left(
1-\varepsilon \right) $\ as coming from the expansion of the
\begin{equation}
\frac{1}{1+\varepsilon }\sim 1-\varepsilon +0\left( 2\right)
\end{equation}
Using this heuristic argument, eq(\ref{approx}) can also be thought of as
\begin{equation}
F\left( \tau _{n}\right) \sim \left( 1+\frac{\tau _{n}-\tau _{n-1}}{\tau
_{n-1}}\right) ^{-1}F\left( \tau _{n-1}\right)
\end{equation}
Doing the same with $F\left( \tau _{n-1}\right) $, one gets by iteration,
the following expression,
\begin{equation}
F\left( \tau _{m}\right) \sim F\left( \tau _{1}\right)
\prod_{j=1}^{m-1}\left( 1+\frac{\tau _{j+1}-\tau _{j}}{\tau _{j}}\right)
^{-1}
\end{equation}
Note also that the $d_{a}$ dependence is in the prefactor of the second term
of the right hand of eq(\ref{ddepend}) allows to cover all possible layer
configurations. For example, in the particular case where $d_{b}-d_{a}=0$ $%
\forall a,b=1,...,n$, the masses of the gauge field components are all of
them zero and the $\mathrm{A}_{\overline{w}}=\mathrm{b}_{\overline{w}%
}^{a}\left( w,\overline{w}\right) \mathbf{\pi }_{a}+\mathrm{b}_{\overline{w}%
}^{ba}$ $\mathbf{u}_{ba}$ gauge field is then in the adjoint representation
of $U\left( n\right) $. The effective field theory one gets is a non
commutative non abelian $U\left( n\right) $ Chern Simons model in $\left(
1+2\right) $ dimensions with level $k$. For the other special case where $%
d_{b}-d_{a}>>>0$ $\forall a\neq b=1,...,n$, the $\mathrm{A}_{\overline{w}}$
gauge field reduces to the massless components $\mathrm{b}_{\overline{w}%
}^{a}\left( w,\overline{w}\right) \mathbf{\pi }_{a}$ valued in the $U\left(
1\right) ^{n}$ Cartan subgroup. Due to their masses, the other gauge
components have heavy dynamics and so decouple.

\subsubsection{NC $U\left( n\right) $ effective field model}

\qquad Using the partition relation of the identity in terms of the
projectors, $\mathrm{I}_{n}=\sum_{a=1}^{n}\mathbf{\pi }_{a}$, one can
express the generalized Susskind map for the case of a $U\left( n\right) $
configuration for the layers, with filling factor $1/k$,
\begin{equation}
z\mathrm{I}=w\mathrm{I}+\theta \mathrm{A,}
\end{equation}
where now $\mathrm{A}_{\overline{w}}$ is in the adjoint of $U\left( n\right)
$. Following the same lines as in subsection 4.2, the NC Chern Simons
effective field model for the system of coincident layers, $\left|
d_{b}-d_{a}\right| <<<1$, with filling factor $\frac{1}{k}$, is given by the
following $U(1)^{n}$ NC Chern Simons theory;
\begin{eqnarray}
S[\mathrm{A}_{\mu }] &=&\frac{k}{4\pi }\int d^{3}y\text{ }\epsilon ^{\mu \nu
\rho }\mathrm{Tr}[\mathrm{\partial }_{\mu }\mathrm{A}_{\nu }\mathrm{\ast A}%
_{\rho }\mathrm{+}\frac{2i}{3}\mathrm{A}_{\mu }\mathrm{\ast A}_{\nu }\mathrm{%
\ast A}_{\rho }-\frac{2}{k}\mathrm{\partial }_{\mu }\mathrm{A}_{\nu }\mathrm{%
\ast }\mathcal{A}_{\rho }]  \notag \\
&&-\frac{1}{2\pi }\int d^{3}y\text{ }\epsilon ^{\mu \nu \rho }\mathrm{Tr}%
[\left( \mathrm{\partial }_{\mu }\mathrm{A}_{\nu }\mathrm{\ast }\mathcal{A}%
_{\rho }\right)
\end{eqnarray}
To get the free filling factor $\mathrm{\nu }$ from the above field action,
it is enough to remember the expression $\theta \sim 1/\left( \mathrm{\nu }%
\mathcal{B}_{ex}\right) $ which shows that the NC parameter depends on $%
\mathrm{\nu }$\ ; but also on the magnitude of the external magnetic field.
In the limit $\mathcal{B}_{ex}$ very large and so $\theta $ small, one can
expand the previous action in powers of $\left( 1/\mathcal{B}_{ex}\right) $.
At the leading order, the above action reduces to the usual commutative non
abelian $U\left( n\right) $ Chern Simons model and so $\mathrm{\nu =n/k}$.
This result is in fact valid to all orders in $\left( 1/\mathcal{B}%
_{ex}\right) $ as in the Susskind NC $U\left( 1\right) $ model with filling
factor $\mathrm{1/k}$. Note that a more complete derivation for the filling
fraction is also possible; it suffices to note that the total number of
particles $\mathcal{N}_{e}$ of the whole system is a conserved quantity,
which in the present field realization reads as,
\begin{equation}
\mathcal{N}_{e}=\int_{\cup \mathcal{L}_{a}}d^{2}y\text{ }\mathcal{J}_{nc}^{0}%
\text{,}
\end{equation}
where now $\mathcal{J}_{nc}^{0}$ is the density of a conserved current $%
\mathcal{J}_{nc}^{\mu }$ on the NC space. Since the conserved current
associated to $\mathcal{N}_{\phi }$ is just $\mathcal{J}_{nc}^{\mu }$ times
the filling fraction $\mathrm{\nu }$, one sees that $\mathrm{\nu }$ appears
as a global factor which does not depend on the way the calculations are
handled.

\subsubsection{NC $U\left( n\right) $ Matrix model for multilayer states}

\qquad The analysis we\ have developed so far applies as well to get the
matrix formulation of the generalization of the Susskind-Polychronakos
regularized matrix model originally obtained for the case of single layer.
As $U\left( n\right) $ gauge symmetry requires that all layers should be
similar; that is with same filling factor $1/k$, it follows that the numbers
$N_{a}$ of particles on $\mathcal{L}_{a}$ should, at any time, be the same;
i.e $N_{1}=...=N_{n}$, and so the total number is $\mathcal{N}=nN$. Keeping
this feature in mind, \ let us proceed as if the numbers $N_{a}$ were
different; in the end we shall set the condition\ $\mathcal{N}=nN$.

Before going ahead recall that in the original Susskind-Polychronakos matrix
model, corresponding to $n=1$ in our present study, the coordinates $%
z_{\alpha }\left( t\right) $, of system of $M$ electrons moving on the plane
are thought of as described by a $M\times M$ matrix field $Z\left( t\right) $
in the adjoint of\ $U\left( M\right) $, with an action given by eq(\ref
{matrixsp}). In the case of the FQH multilayer system we have been
considering here, the diagonal coordinates $z_{\alpha _{aa}}\left( t\right) $
are represented by $N_{a}\times N_{a}$ square matrix fields $Z_{aa}\left(
t\right) $\ while the $z_{\alpha _{ab}}\left( t\right) $ non diagonal one
are represented by $N_{a}\times N_{b}$ \ ``\ rectangular ''\ matrix fields $%
Z_{ab}\left( t\right) $.
\begin{eqnarray}
z_{\alpha _{aa}} &\rightarrow &Z_{aa};\quad a=b  \notag \\
z_{\alpha _{ab}} &\rightarrow &Z_{ab};\quad a\neq b
\end{eqnarray}
The matrix field of the $U\left( n\right) $ model for multilayers is then
given by the $\mathcal{N}\times \mathcal{N}$ matrix $Z\left( t\right) $
transforming in the adjoint of $U\left( \mathcal{N}\right) $ and may be
expanded in different ways by using either the $\mathbf{C}^{\mathcal{N}}$
basis or the $\mathbf{C}^{n}$ one. In the second case, the expansion reads
as,
\begin{equation}
Z\left( t\right) =\sum_{a,b=1}^{n}Z_{ab}|a><b|;
\end{equation}
where $\mathcal{N}=N_{1}+...+N_{n}$ is the total number of particles of $%
\left\{ \cup _{a}\mathcal{L}_{a}\right\} $. Each component $Z_{ab}$ is in
the bifundamental $\left( \mathbf{N}_{a},\overline{\mathbf{N}}_{b}\right) $
of the $U\left( N_{a}\right) \times U\left( N_{b}\right) $ group. Along with
this matrix field and in addition to the auxiliary Lagrange field $\mathrm{A}
$ of similar algebraic nature as $Z$, one has also a generalized
Polychronakos field $\Psi $ in the fundamental of $U\left( \mathcal{N}%
\right) $, which can also be expanded as,
\begin{equation}
\Psi =\sum_{a=1}^{n}\Psi _{a}|a>;
\end{equation}
where each component $\Psi _{a}$ is a polychronakos type field in the
fundamental of $U\left( N_{a}\right) $. The action of the regularized matrix
model for multilayers generalizing the Susskind Polychronakos one reads then
as,
\begin{eqnarray}
S\left[ Z,\Psi ,A\right] &=&{\frac{k}{4\theta }}\int dt\mathrm{Tr}_{U\left(
\mathcal{N}\right) }\left( i{\bar{Z}}DZ-\omega Z{\bar{Z}}\right) +\ hc
\notag \\
&&+\frac{i}{2}\int dt\Psi ^{\dagger }D\Psi +\frac{k}{2}\int dt\mathrm{Tr}%
_{U\left( \mathcal{N}\right) }\mathrm{A}+\ hc,
\end{eqnarray}
where $DZ=\partial _{t}Z+i\left[ A,Z\right] $ and $D\Psi =\left( \partial
_{t}+iA\right) \Psi $. Eliminating the $\mathrm{A}$ auxiliary field, one
gets the following constraint eq
\begin{equation}
\left[ Z,{\bar{Z}}\right] +\frac{2i\theta }{k}\Psi ^{\dagger }\Psi =\mathrm{I%
}_{\mathcal{N}}
\end{equation}
Taking the trace $\mathrm{Tr}_{U\left( \mathcal{N}\right) }$ of this
relation, one obtains $\mathrm{Tr}_{U\left( \mathcal{N}\right) }\left[ \frac{%
2i\theta }{k}\Psi ^{\dagger }\Psi \right] =\mathcal{N}$. Since $\mathcal{N}=%
\mathrm{n}N$ and as $N=\mathcal{N}_{\phi }/\mathrm{k}$, one discovers that
the filling factor of the $U\left( \mathrm{n}\right) $ regularized matrix
model is $n/k$. Moreover knowing that quantum corrections induce a shift of
the level $\mathrm{k}$ by one, we end with the following quantum formula,
\begin{equation}
\mathrm{\nu }=\frac{\mathrm{n}}{\mathrm{k}+1}
\end{equation}
With this relation, we conclude the multilayer representation for FQH states
with rational filling factors and turn now to describe the hierarchical ones.

\section{Hierarchical Buildings}

\qquad FQH states with filling factor that are not of Laughlin type exist
also for the case of a single layer $\mathcal{L}$; that is a FQH system with
a unique $D2$ brane. This is the case for instance for the so called
hierarchical states falling in Jain sequence \cite{jainseq}
\begin{equation}
\nu _{n}^{\left( Jain\right) }=\frac{n}{2np\pm 1}
\end{equation}
and Haldane series \cite{halseri},
\begin{equation}
\nu _{n}^{\left( Hal\right) }=\frac{1}{p_{1}-\frac{1}{p_{2}-...\frac{1}{p_{n}%
}}}
\end{equation}
In the picture; one layer $\mathcal{L}$ $\longleftrightarrow $ one $D2$
brane, where one has only one $U\left( 1\right) $ gauge field, it seems a
priori not obvious to describe such kind of states using Chern Simons
effective field model, which require rather a kind of non abelian symmetry
as it has been done in \cite{14}. There, one is lead to impose extra
constraint eqs to recover the Wen-Zee effective field model for hierarchy
\cite{22}.\ In this section we want to develop an other issue which has the
advantage of being natural, based on a powerful mathematical structure and
does not need imposing constraint eqs. The main idea of this method is based
on having the correspondences; one layer $\mathcal{L}$ $\longleftrightarrow $
one $D2$ brane, and the $\QTR{sl}{LL}$ Landau Levels as a fiber bundles
\textsl{F} above the $\mathcal{D}2$ brane which plays the role of a base
manifold as exposed in subsection 2.2; see figures 2 and 3.

In this picture, hierarchical states with rational filling factors are
described by a Chern Simons effective field model on a fiber bundle \textsl{F%
}$\left( B\right) $ whose base $B$ is identified with the layer $\mathcal{L}$
( $\mathcal{D}2$ brane ) and fiber \textsl{F} given by a vector space
\textsl{V}$,$ whose vector basis are associated with the Landau levels $%
\left\{ \QTR{sl}{LL}_{i}\right\} $\textsl{. }Before giving details, let us
first review some general aspects of FQH states, belonging to Jain and
Haldane hierarchies, that are useful for our later analysis.

\subsection{Effective Field Model}

\qquad In the Wen-Zee effective field theory approach on a single layer,
hierarchical FQH states with rational filling factors are described by a $%
(2+1)$ dimension gauge system of \ several $U\left( 1\right) $ coupled
Chern-Simons fields $\mathrm{b}_{\nu }^{I}$ carrying fractional electric
charge. In this approach, the usual Laughlin state with $\nu _{L}=1/k$ is
viewed as just the leading term $\nu _{L}=\nu \left( n=1\right) $ of a
series $\nu \left( n\right) $, $n=1,2,...$, of FQH states for which general
rational values of $\nu $ appears starting from the level two of the
hierarchy. In this picture, rational values of $\nu \left( n\right) $ emerge
in a natural way and are interpreted as due to successive particle
condensations leading to FQH states occupying several lowest Landau levels
\textsl{LL}$_{i}$. At a generic level $n$, that is for the $n-th$ Landau
level, Wen-Zee model is based on the following action with $n$ Chern Simons
fields coupled as follows \cite{22,23},
\begin{eqnarray}
S\left[ \mathrm{b},\mathcal{A}^{ex},\mathrm{C}_{\mu }^{ex}\right] &=&\frac{1%
}{4\pi }\int d^{3}y\ \epsilon ^{\mu \nu \rho }\mathrm{G}_{ij}\partial _{\mu }%
\mathrm{b}_{\nu }^{i}\mathrm{b}_{\rho }^{j}  \label{wen10} \\
&&-\frac{1}{2\pi }\int d^{3}y\ \left( \epsilon ^{\mu \nu \rho }\mathrm{q}%
_{i}\partial _{\mu }\mathrm{b}_{\nu }^{i}\mathcal{A}_{\sigma }^{ex}+l_{i}%
\mathrm{b}_{\mu }^{i}\mathrm{C}_{ex}^{\mu }\right) ,  \notag
\end{eqnarray}
where $l_{i}$ is an integer and $\mathrm{C}_{ex}^{\mu }$ is an external
source which has been interpreted in \cite{8,14} as the gauge potential
associated with the magnetic $D6$ brane. We will not consider this term here
and so ignore about it.

In the above functional, $\mathrm{G}_{ij}$ is a $n\times n$ matrix, with
specific integer entries, mainly defining the underlying $U\left( 1\right)
^{n}$\ symmetry group of this action and $\mathrm{q}_{i}$ is the
electromagnetic charge vector. Recall in passing that, in addition to the
electomagnetic invariance $U_{em}\left( 1\right) $, this action has an extra
manifest $U\left( 1\right) ^{n}$\ symmetry acting as
\begin{equation}
\mathrm{b}_{\mu }^{i\prime }=\mathrm{b}_{\mu }^{i}+\mathrm{\partial }_{\mu }%
\mathrm{\lambda }^{i};\quad i=1,...,n,
\end{equation}
where $\mathrm{\lambda }^{i}$ are $n$ gauge parameters.\ Invariance under
these transformations is mainly ensured by the antisymmetric property of $%
\epsilon ^{\mu \nu \rho }$ tensor and the conservation of the external
source $\partial _{\mu }\mathrm{C}_{ex}^{\mu }=0$. Note also that $\mathrm{G}%
_{ij}$ and $\mathrm{q}_{i}$\ are interpreted in Wen-Zee model as order
parameters that classify the various FQH states. Their link with the filling
factor is given by the relation,
\begin{equation}
\mathrm{\nu }=\mathrm{q}_{i}\mathrm{G}_{ij}^{-1}\mathrm{q}_{j},
\end{equation}
allowing to distinguish between the various hierarchies. At level $n=2$ for
instance, the Jain and Haldane order parameters $\mathrm{G}_{ij}$ and $%
\mathrm{q}_{i}$ are different; the first ones are given by,
\begin{eqnarray}
\mathrm{G}_{ij} &=&\left(
\begin{array}{cc}
2p+1 & 2p \\
2p & 2p+1
\end{array}
\right) ,  \notag \\
\mathrm{q}_{i} &=&\left( 1,1\right) ,\quad \nu _{2}^{\left( Jain\right) }=%
\frac{2}{4p+1},
\end{eqnarray}
with $p$ an integer, while for Haldane model we have
\begin{eqnarray}
\mathrm{G}_{ij} &=&\left(
\begin{array}{cc}
p_{1} & -1 \\
-1 & p_{2}
\end{array}
\right) ,  \notag \\
\mathrm{q}_{i} &=&\left( 1,0\right) ,\quad \nu _{2}^{\left( Hald\right) }=%
\frac{p_{2}}{p_{1}p_{2}-1}
\end{eqnarray}
where $p_{1}$ odd integer and $p_{2}$\ is even. One of the interesting
feature of the Wen-Zee effective field theory eq(\ref{wen10}), is that it
reduces to the usual CS effective field theory of the Laughlin state with
filling fraction $\nu _{L}$ given by the fraction $\mathrm{G}%
_{11}^{-1}=1/\left( 2p+1\right) $. An other property, is that this effective
field model allows the description of FQH states with rational filling
factor as shown on eqs(5.6-7) for the specific case of level two. In what
follows, we want to show that, in general, there is one to one
correspondence between the level of the hierarchy and the number of occupied
Landau levels $\left\{ \QTR{sl}{LL}_{i}\right\} $ of the layer $\mathcal{L}$%
. \ This property is at the basis of fiber bundle description we have
described in subsection 2.2.

\subsection{More on Jain Hierarchy}

\qquad Jain's FQH states concerns a FQH layer involving the $n$ lowest
Landau levels. Its filling factor $\mathrm{\nu }_{n}\left( p\right) =\frac{n%
}{2np+1}$ follows from a hierarchical matrix $\mathrm{G}=\mathrm{I}_{n\times
n}+2p\mathrm{C}$\textrm{\ }
\begin{equation}
\mathrm{\nu }_{n}\left( p\right) =\mathrm{q}^{T}\mathrm{G.q}
\end{equation}
where $\mathrm{C}$ is a $n\times n$ matrix with all elements equal to one
and where $\mathrm{q}$ is the electric $n$-vector charge equal to\textrm{\ }$%
\left( 1,...,1\right) $. From the structure of the Jain hierarchical matrix $%
\mathrm{G}$, which we prefer to decompose as
\begin{equation}
\mathrm{G}=\left( 1+2p\right) \left( \mathrm{I}_{n\times n}+\frac{2p}{2p+1}%
\mathrm{E}\right)
\end{equation}
where now $\mathrm{E}\ $satisfies,
\begin{equation}
\mathrm{E}^{2}=\left( n-1\right) \mathrm{I}+\left( n-2\right) \mathrm{E,}
\end{equation}
one sees that the total filling factor $\mathrm{\nu }_{n}^{\left(
Jain\right) }\left( p\right) =\mathrm{q}^{T}\mathrm{G}^{-1}\mathrm{q}$ may
be split as the sum of two terms, one diagonal $\mathrm{\nu }_{n}^{\left(
d\right) }\left( p\right) $ and the other $\mathrm{\nu }_{n}^{\left(
nd\right) }\left( p\right) $ non diagonal,
\begin{equation}
\mathrm{\nu }_{n}^{\left( Jain\right) }\left( p\right) =\mathrm{\nu }%
_{n}^{\left( d\right) }\left( p\right) +\mathrm{\nu }_{n}^{\left( nd\right)
}\left( p\right) .  \label{split10}
\end{equation}
In absence of interactions between Landau levels; i.e in case where the $%
\mathrm{G}$ matrix is restricted to the form $\mathrm{G}=\left( 1+2p\right)
\mathrm{I}_{n\times n}$, then the free filling factors of the system is
given by the sum over the factors of each $\QTR{sl}{LL}_{i}$. that is,
\begin{equation}
\mathrm{\nu }_{n}^{\left( free\right) }\left( p\right) =\frac{n}{2p+1}
\end{equation}
where $n$ is also the result of the product $\mathrm{q}^{T}\mathrm{q}$. In
presence of interactions, the filling factor gets extra contributions
induced by couplings. To have an explicit expression of these contributions,
let us rewrite the splitting (\ref{split10}) as,
\begin{equation}
\mathrm{\nu }_{n}^{\left( Jain\right) }\left( p\right) =\sum_{i=1}^{n}\nu
_{n}^{\left( ii\right) }+2\sum_{i\leq j=1}^{n}\nu _{n}^{\left( ij\right) },
\end{equation}
where the diagonal term is
\begin{equation*}
\mathrm{\nu }_{n}^{\left( d\right) }\left( p\right) =\sum_{i=1}^{n}\nu
_{n}^{\left( ii\right) }.
\end{equation*}
As the inverse of the hierarchical matrix is $\mathrm{G}^{-1}=\left[
1-2p/\left( 1+2pn\right) \right] I-2p/\left( 1+2pn\right) E$, one can easily
compute the explicit expressions of the diagonal $\nu _{n}^{\left( ii\right)
}$ and non diagonal $\nu _{n}^{\left( ij\right) }$ component terms. We find
\begin{eqnarray}
\nu _{n}^{\left( ii\right) } &=&\frac{1+2p\left( n-1\right) }{1+2pn},  \notag
\\
\nu _{n}^{\left( ij\right) } &=&\frac{-2p}{1+2pn},\quad i<j
\end{eqnarray}
Summing over the $n$ Landau layers \textsl{LL}$_{i}$, one gets
\begin{eqnarray}
\sum_{i=1}^{n}\nu _{n}^{\left( ii\right) } &=&\frac{\left[ 1+2p\left(
n-1\right) \right] n}{1+2pn},  \notag \\
2\sum_{i<j=1}^{n}\nu _{n}^{\left( ij\right) } &=&\frac{-2pn\left( n-1\right)
}{1+2pn},\quad i<j
\end{eqnarray}
Denoting by
\begin{equation}
\widetilde{\nu }_{eff}^{\left( i\right) }=\sum_{j=1}^{n}\nu _{n}^{\left(
ij\right) },  \label{jainef}
\end{equation}
the filling factor of the $i-th$ lowest $\QTR{sl}{LL}_{i}$ of an effective
FQH\ system of $n$ lowest $\left\{ \cup _{i=1}^{n}\QTR{sl}{LL}_{i}\right\} $%
; but now without interactions, and using eqs(5.14), one finds
\begin{equation}
\widetilde{\nu }_{eff}^{\left( i\right) }=\frac{1}{1+2pn}
\end{equation}
As a result Jain series may be viewed as describing $n$ decoupled effective
lowest $\widetilde{\QTR{sl}{LL}}_{i}$ levels with filling factor given by
eq(5.17). Each effective level is of Laughlin type. In Wen-Zee gauge field
approach, the decomposition (\ref{jainef}) is equivalent to diagonalize the $%
\mathrm{G}_{ij}\partial _{\mu }\mathrm{b}_{\nu }^{i}\mathrm{b}_{\rho }^{j}$
quadratic form.

\subsection{More on Haldane Hierarchy}

\qquad Haldane hierarchy concerns a FQH layer with filling factor $\nu
_{n}^{\left( Hal\right) }$; that is not of the Laughlin type. At a generic
level $n$ of the hierarchy, the filling factor $\nu _{n}^{\left( Hal\right)
}=\nu _{n}\left( p_{1},...,p_{n}\right) $,
\begin{equation}
\nu _{n}^{\left( Hald\right) }=\frac{1}{p_{1}-\frac{1}{p_{2}-...\frac{1}{%
p_{n}}}},
\end{equation}
is a series dependent on $n$ integers and exhibits very special features
basically inherited from those of the continuous fraction,
\begin{equation}
\nu _{n}^{\left( Cont\right) }=\left[ p_{1},...,p_{n}\right]
\end{equation}
This series has a quite similar interpretation to the Jain one in the sense
that it describes an effective system of $n$ lowest $\left\{ \cup _{i=1}^{n}%
\QTR{sl}{LL}_{i}\right\} $ of Laughlin type; but now with different
effective filling factors
\begin{eqnarray}
\widetilde{\nu }_{eff}^{\left( i\right) } &=&\sum_{j=1}^{n}\nu _{n}^{\left(
ij\right) },  \notag \\
\widetilde{\nu }_{eff}^{\left( 1\right) } &\neq &\widetilde{\nu }%
_{eff}^{\left( 2\right) }\neq ...\neq \widetilde{\nu }_{eff}^{\left(
n\right) }.
\end{eqnarray}
The point is that the continuous fraction above exhibits very remarkable
features which allows to bring Haldane states into a system quite similar
the Jain one. Indeed, it is not difficult to check that eq(5.18) can be
usually put into the unique expansion form,
\begin{equation}
\nu _{n}^{\left( h\right) }=\sum_{i=0}^{n-1}\frac{1}{1+2q_{i+1}}.
\end{equation}
where the $q_{i+1}$ integers are given by,
\begin{equation}
q_{i+1}=\frac{m_{i}m_{i+1}-1}{2}
\end{equation}
In these relations the $m_{i}$ are odd integers related to the Haldane $%
p_{i} $'s by help of the following recurrent formula as,
\begin{equation}
m_{0}=1,\quad m_{1}=p_{1}
\end{equation}
together with,
\begin{equation}
m_{i}=p_{i}m_{i-1}-m_{i-2};\qquad 2\leq i\leq n
\end{equation}
Since in Haldane hierarchy $p_{i}$ is even for $2\leq i\leq n$, it follows
that all integers $m_{i}$ are odd and so is the $m_{i}m_{i+1}$ product.
Thus, the filling factor $\nu _{n}^{\left( Hald\right) }$ in the Haldane
hierarchy may be thought as the sum over $n$ effective Laughlin type states
with filling factor $\frac{1}{m_{i}m_{i+1}}$; but built one on the top of
the other as in Fiber bundle theory. With this picture in mind, we propose
now to revisit Wen-Zee field model using the geometric fiber bundle
representation.

\section{Wen-Zee Effective Field Model \ revisited}

\qquad Here above we have shown that a single layer hierarchical states at
level $n$ may be thought of as FQH states filling the $n$ leading Landau
levels. Here, we use the result of subsection 2.2 to give interpretation of
the Wen-Zee model in terms of vector fiber bundles. Such analysis will
allows us to give, amongst others, a geometric interpretation to Jain and
Haldane sequences; but also permits us to build their non commutative field
theory extension. Our construction answers some questions raised in the
literature concerning the Wen-Zee effective field model especially the point
regarding the conserved quantities of the model \cite{21}.

\subsection{Fiber Bundles Approach: Commutative Case}

\qquad Wen-Zee approach to fractional quantum Hall hierarchies on a layer $%
\mathcal{L}$ and its non commutative extension may be given a nice geometric
interpretation in terms of vector\ fiber bundles $\QTR{sl}{F}$ on $\mathcal{D%
}2$ brane. We first reconsider the original Wen-Zee effective field model
and then we study its non commutative extension.

\textit{Commutative Case}\newline
The layer $\mathcal{L}$, viewed as a $\mathcal{D}2$ brane, has world volume
parameterized by the $\left\{ y=\left( y^{\mu }\right) \right\} $
coordinates. Fiber bundles $\QTR{sl}{F}\left( \mathcal{L}\right) =\cup _{y}%
\QTR{sl}{F}_{y}\left( \mathcal{L}\right) $ are parameterized by $\left(
\alpha _{i};y\right) $ where $\alpha _{i}$ are some linearly independent
vectors \textit{globally defined} on $\mathcal{L}$. Each fiber $\QTR{sl}{F}%
_{y}\left( \mathcal{L}\right) $ above the $y$ point, which we have denoted
as $\QTR{sl}{V}\left( \mathcal{L}\right) $, is an infinite dimensional
vector space with a canonical basis eq(2.12), and has proper finite $n$
dimension subspaces $\QTR{sl}{V}_{n}\left( \mathcal{L}\right) $ generated by
$\left\{ \alpha _{i};1\leq i\leq n\right\} $\ and satisfying the embedding
(2.14), as a manifestation of the fact that $\left\{ \cup _{i=1}^{m}\QTR{sl}{%
LL}_{i}\right\} \subset \left\{ \cup _{i=1}^{m+1}\QTR{sl}{LL}_{i}\right\} $.

Given a level of the Wen-Zee hierarchy, say $n$, the intersection matrix $%
\alpha _{i}\cdot \alpha _{j}$ of the vector basis of the fiber bundle is
identified with the matrix $\mathrm{G}_{ij}$;
\begin{equation}
\alpha _{i}\cdot \alpha _{j}=\mathrm{G}_{ij},\quad 1\leq i\leq n
\label{bundle10}
\end{equation}
where $\mathrm{G}_{ij}$ is exactly the matrix appearing in the fields model
( \ref{wen10}). To get the point, let us illustrate the analysis on the
familiar Haldane and Jain examples.

\subsubsection{Haldane bundle}

\qquad Since in Haldane hierarchy, the intersection matrix $\mathrm{G}_{ij}$
depends on $n$ integers $p_{1}$,$...$, $p_{n}$, we shall characterize the
Haldane bundle by these quantum numbers. At a generic level $n$, the
corresponding fiber $\QTR{sl}{V}_{n}\left( \mathcal{L}\right) $ will be
denoted as
\begin{equation}
\QTR{sl}{V}_{n}\left( \mathcal{L}\right) \equiv \QTR{sl}{V}_{n}^{\left\{
p_{1},...,p_{n}\right\} }\left( \mathcal{L}\right) ,
\end{equation}
This bundle can be obtained explicitly by solving eq(\ref{bundle10}) in
terms of the canonical basis eq(2.12). To illustrate further the method, let
us construct the two first leading fibers and then give the general result.

\begin{itemize}
\item  \textsl{Laughlin Fiber}\textit{: }$\QTR{sl}{V}_{1}^{\left\{
p_{1}\right\} }\left( \mathcal{L}\right) $\textit{\ }
\end{itemize}

Since $\mathrm{G}_{11}=p_{1}$, the unique vector $\alpha _{1}$ may solved in
terms of the vector basis (2.12) as follows,
\begin{equation}
\alpha _{1}=\eta \left( \mathbf{e}_{1}+\mathbf{e}_{2}+...+\mathbf{e}%
_{p_{1}}\right)
\end{equation}
where $\eta $\ is a number. Taking the norm of this vector; i.e
\begin{equation}
\alpha _{1}\cdot \alpha _{1}=p_{1}
\end{equation}
one sees that $\eta $ may be set to $\pm 1$;. we will choose $\eta =1$. The
choice of the plus sign in front of each $\mathbf{e}_{i}$ is due to the
hypothesis requiring a complete symmetry under permutations between the $%
\mathbf{e}_{i}$ canonical vectors. Note in passing, though we suspect that
the $\mathbf{e}_{i}$'s might have some link with the number of quantum
fluxes per surface unit, namely $p_{1}$; we have not yet succeeded to figure
out convincing arguments valid at all levels of the hierarchy.

\begin{itemize}
\item  \textsl{Fiber}\textit{: }$\QTR{sl}{V}_{2}^{\left\{
p_{1},p_{2}\right\} }\left( \mathcal{L}\right) $\textit{\ }
\end{itemize}

As we have said, this fiber describes the level two of the Haldane
hierarchy. Using the result (6.3) and putting \ back into eq(6.1), one can
solve the expression of $\alpha _{2}$. This solution, which can be chosen
as,
\begin{equation}
\alpha _{2}=-\mathbf{e}_{p_{1}}-\left( \mathbf{e}_{p_{1}+1}+...+\mathbf{e}%
_{p_{2}-1}\right) .  \label{alpha2}
\end{equation}
is however not unique since the role of $\mathbf{e}_{p_{1}}$ can be played
by any vector $\mathbf{e}_{k}$, with $1\leq k\leq p_{1}$. In other words,
the above expression has the manifest symmetry
\begin{equation}
\alpha _{2}\rightarrow \alpha _{2}+\left( \mathbf{e}_{p_{1}}-\mathbf{e}%
_{k}\right) ,1\leq k\leq p_{1}
\end{equation}
preserving the norm $\left\| \alpha _{2}\right\| ^{2}=p_{2}$ and leaving $%
\alpha _{1}\cdot \alpha _{2}=-1$. Thus, the choice eq(\ref{alpha2})
corresponds to a fixing of the symmetry $\alpha _{2}\rightarrow \alpha
_{2}+\left( \mathbf{e}_{p_{1}}-\mathbf{e}_{k}\right) $ of the Haldane matrix.

The gauge field $\mathbf{A}_{\mu }$ describing the effective field dynamics
of the electrons moving in this bundle depends on both the coordinate point $%
y$ and the two directions $\alpha _{1}$ and $\alpha _{2}$ of the fiber $%
\QTR{sl}{V}_{2}^{\left\{ p_{1},p_{2}\right\} }$. In other words, the gauge
field $\mathbf{A}_{\mu }$ should be thought of as
\begin{equation}
\mathbf{A}_{\mu }\left( y\right) =\alpha _{1}\mathrm{b}_{\mu }^{1}\left(
y\right) +\alpha _{2}\mathrm{b}_{\mu }^{2}\left( y\right)
\end{equation}
overcoming by the occasion the apparent difficulty figuring on the original
Wen-Zee effective field model. With this field realization, the Wen-Zee idea
follows naturally from the following universal action
\begin{equation}
S\left[ \mathbf{A}_{\mu },\mathcal{A}_{\mu }^{ex}\right] =\frac{\mathrm{1}}{%
4\pi }\int d^{3}y\varepsilon ^{\mu \nu \sigma }\left[ \mathbf{A}_{\mu
}\partial _{\nu }\mathbf{A}_{\sigma }-2\mathbf{q}.\mathbf{A}_{\mu }\partial
_{\nu }\mathcal{A}_{\sigma }^{ex}\right] ,
\end{equation}
where the $\mathbf{q}$ vector charge is such that $\mathbf{q.}\alpha _{1}=1$
and $\mathbf{q.}\alpha _{2}=0$.

\begin{itemize}
\item  \textsl{Generic Levels}
\end{itemize}

The general result for the realization of the $\alpha _{i}$'s in terms of
the canonical vectors eq(2.12) is,

\begin{equation}
\alpha _{r}=\left( -\right) ^{r+1}\left( \mathbf{e}_{p_{r-1}}+%
\sum_{r=1}^{p_{r}-1}\mathbf{e}_{p_{r+j}}\right) .
\end{equation}
Here also the $\alpha _{i}$'s are invariant under permutations of the
canonical vectors corresponding to each \textsl{LL}$_{i}$ and may be
expressed into different; but equivalent, ways as noted before. In this case
the gauge field bundle decomposition leading to the Wen-Zee effective field
model at a level $n$ is given by,
\begin{equation}
\mathbf{A}_{\mu }\left( y\right) =\sum_{i=1}^{n}\alpha _{i}\mathrm{b}_{\mu
}^{i}\left( y\right) ,
\end{equation}
while the $\mathbf{q}$ vector charge is such that $\mathbf{q.}\alpha _{1}=1$
and $\mathbf{q.}\alpha _{i}=0$ for $i\geq 2$.

\subsubsection{Jain Bundle}

\qquad The intersection matrix $\mathrm{G}_{ij}$ of Jain hierarchy depends
on two quantum numbers $n$ and $p$ as shown on eq(5.6). At a generic level $%
n $, the corresponding Jain fiber $\QTR{sl}{V}_{n}\left( \mathcal{L}\right) $
denoted as
\begin{equation}
\QTR{sl}{V}_{n}\left( \mathcal{L}\right) \equiv \QTR{sl}{V}_{n}^{\left\{
p\right\} }\left( \mathcal{L}\right) ,
\end{equation}
is obtained by solving eq(6.1). To illustrate the method, note first that $%
\QTR{sl}{V}_{1}^{\left\{ 2p+1\right\} }\left( \mathcal{L}\right) $ is the
same as in Haldane hierarchy eq(6.3). For the generic case where $n$ and $p$
are arbitrary positive integers, a solution for the Jain matrix, $\mathrm{G=I%
}_{n\times n}\mathrm{+}\left( \mathrm{p}_{1}-1\right) \mathrm{C}$, is given
by the following $\alpha _{i}$ vectors
\begin{equation}
\alpha _{i}=\beta _{0}+\mathbf{e}_{i};\qquad 1\leq i\leq n
\end{equation}
where $\beta _{0}$ is a fixed vector of norm $\left\| \beta _{0}\right\|
=\left( p_{1}-1\right) $, realized as,
\begin{equation}
\beta _{0}=\sum_{k=1}^{p_{1}-1}\mathbf{e}_{-k}
\end{equation}
The $\alpha _{i}$ vectors have norms $\left\| \alpha _{i}\right\| ^{2}=p_{1}$%
\ and their intersection matrix $\mathrm{G}=\left( \mathbf{\alpha }_{i}\cdot
\mathbf{\alpha }_{j}\right) $\ for $i\neq j$ is equal to\textrm{\ }$\left(
\mathrm{p}_{1}-1\right) \mathrm{C}$ and for $i=j$ is indeed $\mathrm{I}%
_{n\times n}\mathrm{+}\left( \mathrm{p}_{1}-1\right) \mathrm{C}$. The
Wen-Zee effective theory follows naturally from the action (6.8) by using
eqs(6.10) and the relation (6.1).

\subsection{Non Commutative Case}

\qquad Here we describe the non commutative extension of Wen-Zee effective
field model for hierarchical FQH states by using the Fiber bundle approach.
This construction goes beyond the Susskind NC model and deals with FQH
states, living in more than one lowest Landau level; that is states which
are not of Laughlin type. This fiber bundle construction seems to us as the
more natural framework to study NC extension of Wen-Zee effective field
model on a single layer. To get the key point of our procedure, we have
judged useful to first derive what is the classical description behind
Wen-Zee model; that is the classical description where one is dealing with a
finite number of classical particles moving on plane. Then we turn to study
the non commutative extension of Wen-Zee field and regularized matrix models

\subsubsection{Classical analysis}

\qquad Since the order $n$ of the hierarchy is also the number of lowest
Landau level $\left\{ \QTR{sl}{LL}_{i}\right\} $ of the FQH layer $\mathcal{L%
}$ that are occupied by electrons, let $N_{i};\quad i=1,...,n$,\ denote the
number of electrons, at a given time, in $\QTR{sl}{LL}_{i}$ and $\mathcal{N}%
_{e}$ the total number of electrons moving in the layer. Of course these $%
N_{i}$ numbers are in general not conserved quantities; only their sum $%
\mathcal{N}_{e}$ which does,
\begin{eqnarray}
\mathcal{N}_{e} &=&N_{1}+...+N_{n} \\
\frac{d\mathcal{N}_{e}}{dt} &=&0;\quad \mathcal{N}_{e}=\nu \mathcal{N}_{\phi
},  \notag
\end{eqnarray}
where $\mathcal{N}_{\phi }$ is as used before. Let also $x_{I_{i}}^{\mu
}\left( t\right) $; $\mu =1,2$, $I_{i}=1,...,N_{i}$; $i=1,...,n$, denote the
space positions of the electrons moving in $\mathcal{L}$ and $v_{I_{i}}^{\mu
}\left( t\right) =\partial _{t}x_{I_{i}}^{\mu }\left( t\right) $ their
velocities. With help of the fiber bundle description we have introduced
above, these variables can be put into the following compact form,
\begin{eqnarray}
\mathbf{x}_{I}^{\mu }\left( t\right) &=&\sum_{i=1}^{n}\alpha
_{i}x_{I_{i}}^{\mu }\left( t\right) ;\quad I=1,...,\mathcal{N}_{e}  \notag \\
\mathbf{v}_{I}^{\mu }\left( t\right) &=&\sum_{i=1}^{n}\alpha
_{i}v_{I_{i}}^{\mu }\left( t\right)
\end{eqnarray}
Taking n=1, one discovers the usual simplest case associated with Susskind
and Susskind-Polykronakos models \cite{1,5}. Like in the standard case, the
classical dynamics of this system of charged particles in presence of a
strong enough external magnetic field $\mathcal{B}_{ex}$ is mainly governed
by the following action,
\begin{equation}
S[\mathbf{x,v}]=\frac{e\mathcal{B}_{ex}}{2}\int dt{\ \varepsilon _{0\mu \nu }%
}\sum_{I=1}^{\mathcal{N}_{e}}\mathbf{v}_{I}^{\mu }\mathbf{x}_{I}^{\nu
}\left( t\right) ,  \label{action}
\end{equation}
or equivalently by substituting $\mathbf{v}_{I}^{\mu }$ and $\mathbf{x}%
_{I}^{\nu }$ by their initial expressions,
\begin{equation}
S\left[ x,v\right] =\frac{e\mathcal{B}_{ex}}{2}\int dt{\ \varepsilon _{0\mu
\nu }}\left( \sum_{i,j=1}^{n}\mathrm{G}_{ij}\text{ }\partial
_{t}x_{I_{i}}^{\mu }x_{I_{j}}^{\nu }\right) ,  \label{action2}
\end{equation}
where $\mathrm{G}_{ij}$ is as in eq(6.1). In the thermodynamic limit where $%
\mathcal{N}_{e}$ goes to infinity, the one dimensional vectors $\mathbf{x}%
_{I}^{\mu }\left( t\right) $; $\mu =1,2$, are maps to a $\left( 1+2\right) $%
-dimensional vector fields $\mathbf{X}_{I}^{\mu }\left( t,y^{1},y^{2}\right)
=\alpha _{i}X_{I_{i}}^{\mu }\left( t,y^{1},y^{2}\right) $ and the condition
of conservation of the total number of particles (6.14) is mapped to a
condition of existence of a $\left( 1+2\right) $ conserved current $\mathcal{%
J}^{\mu }$,
\begin{equation}
\partial _{\mu }\mathcal{J}^{\mu }=\frac{\partial \mathrm{\rho }_{0}}{%
\partial t}+\partial _{i}\mathcal{J}^{i}=0
\end{equation}
The current density $\mathcal{J}^{0}=\mathrm{\rho }$ is just the density of
electrons per surface unit of the layer $\mathcal{L}$ and so
\begin{equation}
\mathcal{N}_{e}=\int_{\mathcal{L}}d^{2}y.\mathrm{\rho }_{0}\mathrm{;\quad }%
\frac{d\mathcal{N}_{e}}{dt}=\int_{\mathcal{L}}d^{2}y\frac{\partial \mathrm{%
\rho }_{0}}{\partial t}.
\end{equation}
There are different ways to handle such relation; a way to do is to take the
simple scenario where $\mathrm{\rho }_{0}$ is time independent and the flux
of particles is conserved; i.e
\begin{equation}
\frac{\partial \mathrm{\rho }_{0}}{\partial t}=\partial _{i}\mathcal{J}^{i}=0
\end{equation}
A furthermore simple choice is to suppose that, in addition to the above
relations, the density $\mathrm{\rho }_{0}$ is uniform as well like in
Susskind formulation for Laughlin states;
\begin{equation}
\frac{\partial \mathrm{\rho }_{0}}{\partial t}=\frac{\partial \mathrm{\rho }%
_{0}}{\partial y^{i}}=0,
\end{equation}
This is the choice we will adopt in this discussion. Note that under
diffeomorphisms of the plane, i.e $y^{\prime }=y^{\prime }\left( y\right) $,
the total number $\mathcal{N}_{e}$ of electrons in $\left\{ \left( \cup _{i}%
\QTR{sl}{LL}_{i};\mathcal{L}\right) \right\} $\ is invariant and so one has
\begin{equation}
\mathrm{\rho }_{0}=\mathrm{\rho }^{\prime }\left| \frac{\partial ^{2}\left(
y^{\prime }\right) }{\partial ^{2}\left( y\right) }\right| ^{2}.
\end{equation}
This relation implies, amongst others, that in the large $\mathcal{N}_{e}$
limit, the discrete sum and integral over the coordinates of $\mathcal{L}$
are related as

\begin{equation}
\sum_{I=1}^{\mathcal{N}_{e}}...\longleftrightarrow \int_{\mathcal{L}}d^{2}y.%
\mathrm{\rho }_{0}\mathrm{,}
\end{equation}
Note in passing that like in multilayer system, the total number $\mathcal{N}%
_{e}$ of particles in $\left\{ \cup _{i}\QTR{sl}{LL}_{i}\right\} $, at any
time, should be thought of as given by a sum over the $N_{i}$ numbers of
particles in the Landau levels $\QTR{sl}{LL}_{i}$ but also on the $N_{ij}$
numbers of particles traveling between the Landau levels. This property
which goes along with the natural identity,
\begin{equation}
\sum_{I=1}^{\mathcal{N}_{e}}...=\sum_{i=1}^{n}\sum_{I_{i}=1}^{N_{i}}...
\end{equation}
suggests that, at any time, the total density $\mathrm{\rho }_{0}$ is in
fact given a sum over the $\mathrm{\rho }_{0}^{\left( ij\right) }$
densities; $n$ densities of particles $\mathrm{\rho }_{0}^{\left( i\right) }$
concerning the Landau levels $\QTR{sl}{LL}_{i}$; but also densities $\mathrm{%
\rho }_{0}^{\left( ij\right) }$\ concerning those particles traveling
between Landau levels. As such we have;
\begin{equation}
\mathrm{\rho }_{0}=\sum_{i,j=1}^{n}\mathrm{\rho }_{0}^{\left( ij\right) }.
\end{equation}
which may also be expressed, in terms of the $\alpha _{i}$ vector basis of
the fiber bundle, as
\begin{equation}
\mathrm{\rho }_{0}=\sum_{i,j=i}^{n}\alpha _{i}.\alpha _{i}\text{ }\mathrm{%
\eta }_{ij}.
\end{equation}
In the present generalized Susskind analysis, we will not need eqs(6.25-26);
and so forget about them for the moment. Putting the correspondence eq(6.23)
in eq(6.16), one gets the following $\left( 1+2\right) $ dimension field
action
\begin{equation}
S[\mathbf{X,V}]=\frac{e\mathcal{B}_{ex}}{2}\int_{\mathcal{L}}d^{3}y{\ }\text{%
\ }{\mathrm{\rho }}_{0}\text{\textrm{\ \ }}{\varepsilon _{0\mu \nu }}\mathbf{%
V}^{\mu }\mathbf{X}^{\nu },  \label{Suss1}
\end{equation}
This is a constrained system; the conjugate momentum $\Pi _{\mu }$ of the
field variable $\mathbf{X}^{\mu }$, which reads, as
\begin{equation}
\Pi _{\mu }=\frac{e\mathcal{B}_{ex}}{2}{\mathrm{\rho }}_{0}\text{\textrm{\ \
}}{\varepsilon _{0\mu \nu }}\mathbf{X}^{\nu },\quad \mu =0,1,2,
\end{equation}
has a vanishing time component and so the action can also rewritten as
\begin{equation}
S[\mathbf{X,}\Pi ]=-\int_{\mathcal{L\times R}}d^{3}y{\ }\Pi _{\mu }.\mathbf{X%
}^{\mu }
\end{equation}
Since the following change under area preserving diffeomorphisms of the
plane
\begin{equation}
\delta \mathbf{X}^{\nu }=\varepsilon ^{0\alpha \beta }\partial _{\alpha }%
\mathbf{X}^{\nu }\partial _{\beta }\lambda ,
\end{equation}
where $\lambda $\ is the parameter of the group, is a symmetry of the
action, the quantity
\begin{equation}
\mathcal{N}_{\phi }=\int_{\mathcal{L}}d^{2}y{\ }.\varepsilon ^{0\alpha \beta
}\partial _{\beta }\Pi _{\nu }\partial _{\alpha }\mathbf{X}^{\nu
}=\oint_{\Gamma }\Pi _{\nu }\delta \mathbf{X}^{\nu },
\end{equation}
is a constant of motion, that is $d\mathcal{N}_{\phi }/dt=0$ which implies
in turns that
\begin{equation}
\partial _{t}\left( \varepsilon ^{0\alpha \beta }\partial _{\beta }\Pi _{\nu
}\partial _{\alpha }\mathbf{X}^{\nu }\right) =-\partial _{\beta }\left(
j^{\beta }\right) .
\end{equation}
This eq is naturally solved for $j^{\alpha }=\varepsilon ^{0\alpha \beta
}\Pi _{\nu }\partial _{\alpha }\mathbf{X}^{\nu }$. In absence of vortices,
the charge density $\varepsilon ^{0\alpha \beta }\partial _{\beta }\Pi _{\mu
}\partial _{\alpha }\mathbf{X}^{\nu }$ of the quantum number flux can also
be written, up to a multiplicative global factor $e\mathcal{B}_{\phi }{%
\mathrm{\rho }}_{0}$, as
\begin{equation}
\frac{1}{2}{\varepsilon _{0\nu \mu }}\varepsilon ^{0\alpha \beta }\partial
_{\beta }\mathbf{X}^{\mu }\partial _{\alpha }\mathbf{X}^{\nu }=\frac{1}{2}{%
\varepsilon _{0\mu \nu }}\left\{ \mathbf{X}^{\mu },\mathbf{X}^{\nu }\right\}
=1,  \label{const1}
\end{equation}
where $\left\{ f\left( y\right) ,g\left( y\right) \right\} =\varepsilon
^{0\alpha \beta }\partial _{\alpha }f\partial _{\beta }g$ stands for the
usual Poisson bracket. Such constraint eq, which reflects just the link
between the two constant of motion $\mathcal{N}_{\phi }$\ and $\mathcal{N}%
_{e}$,
\begin{equation}
\mathrm{\nu }\mathcal{N}_{\phi }=\mathcal{N}_{e},
\end{equation}
has a very remarkable structure and deep consequences. By using the usual
star product $\ast $ of NC geometry on the Moyal plane, eq(6.33) can be
rewritten in a more suggestive form as follows
\begin{equation}
\mathbf{X}^{\mu }\ast \mathbf{X}^{\nu }-\mathbf{X}^{\nu }\ast \mathbf{X}%
^{\mu }=\theta \varepsilon ^{0\mu \nu }+0\left( \theta ^{2}\right)
\end{equation}
where the non commutative $\theta =\frac{1}{2\pi {\mathrm{\rho }}_{0}}=\frac{%
1}{e\mathcal{B}_{\phi }\mathrm{\nu }}$. Moreover as $\mathbf{X}^{\mu }\ast
\mathbf{X}^{\nu }=\mathrm{G}_{ij}X_{i}^{\mu }\ast X_{j}^{\nu }$, one
discovers that the $\theta $ non commutative parameter is in fact given by
the following trace
\begin{equation}
\mathrm{\theta }=\mathrm{G}_{ij}\mathrm{\theta }^{ij},
\end{equation}
in agreement with ${\mathrm{\rho }}_{0}=\mathrm{G}_{ij}{\mathrm{\eta }}%
^{\left( ij\right) }$eq(6.26), and the Susskind observation regarding
Laughlin states.

As usual, the constraint relation(6.33) can be implemented into the action
(6.27), by introducing a Lagrange parameter $\Lambda _{0}$. The result is
\begin{equation}
S[\mathbf{X,}\Lambda _{0}]=\frac{e{\mathrm{\rho }}_{0}\mathcal{B}_{ex}}{2}%
\int_{\mathcal{L}}d^{3}y{\ .\varepsilon _{0\mu \nu }}\left[ \left( \partial
_{0}\mathbf{X}^{\mu }-\mathrm{\theta }\left\{ \mathbf{X}^{\mu },\Lambda
_{0}\right\} \right) \mathbf{X}^{\nu }+\mathrm{\theta }\varepsilon ^{0\mu
\nu }\Lambda _{0}\right] .  \label{continactio}
\end{equation}
At this level, one may derive directly the analogue of Susskind
Polychronakos matrix model of Laughlin states to the case of hierarchical
ones; let us give this result in our path for the quest for the general
effective field model for FQH states with rational filling factors.

\subsubsection{Matrix model for Hierarchical states}

\qquad Using the correspondence (6.23), and viewing now $\mathbf{X}^{\mu }$
as \ $\mathcal{N}_{e}\times \mathcal{N}_{e}$\ hermitian matrix with an
expansion as in eq(6.10), where the $\mathcal{N}_{e}\times \mathcal{N}_{e}$
component matrices $X_{i}^{\mu }$ have only $N_{i}$ non zero diagonal
elements corresponding to the number of electrons on the Landau levels
\textsl{LL}$_{i}$. For instance the $X_{1}^{\mu }$ matrix is as follows,

\begin{eqnarray}
\left( X_{1}^{\mu }\right) _{II} &\neq &0\text{ \ \ \ for \ \ }1\leq I\leq
N_{1}  \notag \\
\left( X_{1}^{\mu }\right) _{II} &=&0\text{ \ \ \ for \ \ }\left(
N_{1}+1\right) \leq I\leq \mathcal{N}_{e}\text{.}
\end{eqnarray}
More generally, we have for $i=1,...,n$, the following\ matrix
representation
\begin{eqnarray}
\left( X_{i}^{\mu }\right) _{II} &\neq &0\text{ \ \ \ for \ \ }\left(
N_{i-1}+1\right) \leq I\leq N_{i}  \notag \\
\left( X_{1}^{\mu }\right) _{II} &=&0\text{ \ \ \ Otherwise for.}
\end{eqnarray}
The way in which the $X_{i}^{\mu }$'s couple to each others is given by the
explicit form of the $\mathrm{G}_{ij}$ intersection matrix of the fiber
bundle. Setting
\begin{equation}
\mathbf{Z}=\mathbf{X}^{1}+i\mathbf{X}^{2};\quad \overline{\mathbf{Z}}=%
\mathbf{X}^{1}-i\mathbf{X}^{2},
\end{equation}
together with the expansion
\begin{equation}
\mathbf{Z=}\sum_{i=1}^{n}\alpha _{i}\mathbf{Z}_{i};\quad \overline{\mathbf{Z}%
}=\sum_{i=1}^{n}\alpha _{i}\overline{\mathbf{Z}}_{i}
\end{equation}
\ the regularized finite $\mathcal{N}_{e}$ version of the continuous action
eq(\ref{continactio}) reads as
\begin{eqnarray}
S\left[ \mathbf{Z},\mathbf{\Psi },\Lambda \right] &=&{\frac{1}{4\mathrm{\nu }%
\theta }}\int dt\mathrm{Tr}\left( i\mathbf{\bar{Z}}D\mathbf{Z}-\omega
\mathbf{Z\bar{Z}}\right) +hc  \notag \\
&&\frac{i}{2}\int dt\mathbf{\Psi }^{\dagger }D\mathbf{\Psi }+\frac{1}{2%
\mathrm{\nu }}\int dtTrA+\ hc.  \label{discrtactio}
\end{eqnarray}
where $A$ is a time component gauge field playing the role of a Lagrange
matrix field parameter and $D$ is the usual time covariant derivative.
Finally $\mathbf{\Psi }$ is a kind of generalized Polychronakos field
playing the role of a regulator; it has an expansion type eq(6.10).
Variation under the Lagrange parameter leads to the constraint equation,
\begin{equation}
\left[ \mathbf{Z},\mathbf{\bar{Z}}\right] +2i\theta \mathrm{\nu }\mathbf{%
\Psi }^{\dagger }\mathbf{\Psi }=\mathrm{I}_{\mathcal{N}_{e}}
\end{equation}
or equivalently, up on taking the trace and using the fiber bundle
expansion,
\begin{equation}
\mathrm{Tr}\mathbf{\Psi }_{i}^{\dagger }\mathbf{\Psi }_{j}=-i\frac{\mathcal{N%
}_{e}}{2\theta \mathrm{\nu }}\mathrm{G}_{ij}^{-1}\text{.}
\end{equation}

\subsubsection{ NC extension of Wen-Zee field Model}

\qquad Following Susskind idea, the NC extension of the Wen-Zee gauge field
model may be obtained by starting from eq(6.27) and making the change of
variables,
\begin{equation}
\mathbf{X}^{\mu }\left( y\right) =\varepsilon ^{0\mu \nu }\mathbf{H}_{\nu
}\left( y\right) ,\quad \mu =1,2
\end{equation}
where $\mathbf{H}_{\nu }\left( y\right) $ are taken as,
\begin{eqnarray}
\mathbf{H}_{\nu }\left( t,y\right) &=&\sum_{i=0}^{n}\mathbf{\theta }_{i}%
\mathrm{b}_{\nu }^{i}\left( t,y\right) =\mathbf{\theta }_{o}\mathrm{b}_{\nu
}^{0}+\sum_{i=1}^{n}\mathbf{\theta }_{i}\mathrm{b}_{\nu }^{i}  \label{split}
\\
&=&\theta \sum_{i=0}^{n}\alpha _{i}\mathrm{b}_{\nu }^{i}=\theta \mathbf{b}%
_{\nu }.
\end{eqnarray}
In this expansion, $\theta $ is a dimensionful parameter scaling as $\left(
lenght\right) ^{2}$\ and $\mathrm{b}_{\nu }^{i}$ scale are gauge fields.
These are fluctuations around the time independent background $y^{\mu
}=\varepsilon ^{0\mu \nu }\theta \mathrm{b}_{\nu }^{0}$. The missing time
component $\mathrm{b}_{0}^{i}$, will be generated later on from the
constraint eqs. Putting this change in the action (6.27), one gets
\begin{equation}
S[\mathbf{b}]=-\frac{e{\mathrm{\rho }}_{0}\theta ^{2}\mathcal{B}_{ex}}{2}%
\int_{\mathcal{L}}d^{3}y{\ }\text{\ }\varepsilon ^{0\alpha \beta }\text{ }%
\partial _{0}\mathbf{b}_{\alpha }\mathbf{b}_{\beta },
\end{equation}
which reduces to sums over $i$ an $j$ starting from $1$ to $n$,
\begin{equation}
S[\mathrm{b}]=-\frac{e{\mathrm{\rho }}_{0}\theta ^{2}\mathcal{B}_{ex}}{2}%
\int_{\mathcal{L}}d^{3}y{\ .}\varepsilon ^{0\mu \nu }\sum_{i,j=1}^{n}\alpha
_{i}.\alpha _{j}\partial _{0}\mathrm{b}_{\mu }^{i}\mathrm{b}_{\nu }^{j},
\label{suss2}
\end{equation}
where we have used the property $\partial _{0}\mathrm{b}_{\mu }^{0}=0$.
Doing the same thing for the constraint eq(\ref{const1}),
\begin{equation}
\frac{\theta ^{2}}{2}\varepsilon ^{0\mu \sigma }\left\{ \mathbf{b}_{\mu },%
\mathbf{b}_{\sigma }\right\} =\frac{\theta ^{2}}{2}\varepsilon ^{0\mu \sigma
}\sum_{i,j=0}^{n}\alpha _{i}.\alpha _{j}\left\{ \mathrm{b}_{\mu }^{i},%
\mathrm{b}_{\nu }^{j}\right\} =1
\end{equation}
or equivalently by setting $\alpha _{0}.\alpha _{0}=1$ and using the
identity used $\left\{ y^{n},\mathrm{h}\right\} =\varepsilon ^{nm}\partial
_{m}\mathrm{h}$,
\begin{equation}
\varepsilon ^{0\mu \nu }\left( \sum_{i=1}^{n}\alpha _{0}.\alpha _{i}\partial
_{\lbrack \nu }\mathrm{b}_{\mu ]}^{i}-\theta \sum_{i,j=1}^{n}\alpha
_{i}.\alpha _{j}\left\{ \mathrm{b}_{\mu }^{i},\mathrm{b}_{\nu }^{j}\right\}
\right) =0,
\end{equation}
Setting $\mathrm{G}_{0i}=\mathbf{\alpha }_{0}\cdot \mathbf{\alpha }_{i}$,
this constraint eq may be rewritten in a more condensed form as,
\begin{equation}
\varepsilon ^{0\mu \nu }\left( \mathrm{G}_{0i}\partial _{\lbrack \nu }%
\mathrm{b}_{\mu ]}^{i}-\theta \mathrm{G}_{ij}\left\{ \mathrm{b}_{\mu }^{i},%
\mathrm{b}_{\nu }^{j}\right\} \right) =0.
\end{equation}
Note that due to this constraint eq, the original Wen-Zee $n\times n$ matrix
$\mathrm{G}_{ij}$ is now extended to a $\left( n+1\right) \times \left(
n+1\right) $ matrix $\mathrm{S}_{ij}$ of the form; see also eq(2.17),
\begin{equation}
\mathrm{G}_{ij}=\left(
\begin{array}{cc}
\mathrm{1} & \mathrm{G}_{0j} \\
\mathrm{G}_{i0} & \mathrm{G}_{ij}
\end{array}
\right)  \label{kac}
\end{equation}
Injecting this constraint eq into the action (\ref{suss2}) by help of a
Lagrange field multiplier $\Lambda $, we get the following
\begin{eqnarray}
S\left[ \mathrm{b}\right] &=&-\frac{e{\mathrm{\rho }}_{0}\theta ^{2}\mathcal{%
B}_{ex}}{2}\int_{\mathcal{L}}d^{3}y{\ .}\varepsilon ^{0\mu \nu }\left[
\mathrm{G}_{ij}\partial _{0}\mathrm{b}_{\mu }^{i}\mathrm{b}_{\nu }^{j}\right]
\label{suss3} \\
&&+\frac{e{\mathrm{\rho }}_{0}\theta ^{2}\mathcal{B}_{ex}}{2}\int_{\mathcal{L%
}}d^{3}y{\ .}\varepsilon ^{0\mu \nu }\left[ \mathrm{\Lambda }\left( \mathrm{G%
}_{0i}\partial _{\lbrack \nu }\mathrm{b}_{\mu ]}^{i}-\theta \mathrm{G}%
_{ij}\left\{ \mathrm{b}_{\mu }^{i},\mathrm{b}_{\nu }^{j}\right\} \right)
\right] .  \notag
\end{eqnarray}
Then introducing the time component gauge fields $\mathrm{b}_{0}^{i}\left(
y\right) $ by setting
\begin{equation}
\mathrm{\Lambda G}_{0j}=-\mathrm{G}_{ij}\mathrm{b}_{0}^{i}
\end{equation}
or equivalently
\begin{equation}
\mathrm{\Lambda }=-\mathrm{G}_{ij}\mathrm{G}_{j0}^{-1}\mathrm{b}%
_{0}^{i};\quad \mathrm{b}_{0}^{i}=-\mathrm{\Lambda G}_{0j}\mathrm{G}%
_{ji}^{-1}
\end{equation}
one can bring the action (\ref{suss3})\ into the more familiar form,
\begin{equation}
S\left[ \mathrm{b}\right] =\frac{1}{4\pi }\int_{\mathcal{L}}d^{3}y{\ .}%
\varepsilon ^{\mu \nu \sigma }\left[ \mathrm{L}_{ij}\partial _{\mu }\mathrm{b%
}_{\nu }^{i}\mathrm{b}_{\sigma }^{j}-\frac{\theta }{3}\mathrm{C}%
_{ijk}\left\{ \mathrm{b}_{\mu }^{i},\mathrm{b}_{\nu }^{j}\right\} \mathrm{b}%
_{\sigma }^{k}\right] ,
\end{equation}
where\
\begin{eqnarray}
\mathrm{L}_{ij} &=&e\theta \mathcal{B}_{ex}\mathrm{G}_{ij};\qquad i,j=1,...,n
\notag \\
\mathrm{C}_{ijk} &=&\frac{e\theta \mathcal{B}_{ex}}{3}\mathrm{G}_{(ij}%
\mathrm{G}_{k)l}\mathrm{G}_{l0}^{-1}.
\end{eqnarray}
In deriving this relation, we have used the antisymmetry and the
associativity $\left\{ \mathrm{b}_{\mu }^{i},\mathrm{b}_{\nu }^{j}\right\}
\mathrm{b}_{0}^{k}=\mathrm{b}_{\mu }^{i}\left\{ \mathrm{b}_{\nu }^{j},%
\mathrm{b}_{0}^{k}\right\} $ properties of the Poisson braket as well as the
cyclicity identity $\mathrm{C}_{ijk}=\mathrm{C}_{jki}=\mathrm{C}_{kij}$.
Note that the presentation we have given above may be also expressed in a
more condensed form by using fiber bundle vector basis. In this way of
doing, the constraint reduces to
\begin{equation}
\varepsilon ^{0\mu \nu }\left( \mathbf{\partial }_{[\nu }\mathbf{b}_{\mu
]}-\theta \left\{ \mathbf{b}_{\mu },\mathbf{b}_{\nu }\right\} \right) =0,
\end{equation}
where we have set,
\begin{equation}
\mathbf{\partial }_{\nu }\mathbf{=\gamma }\text{ }\partial _{\nu },
\end{equation}
and where the $\mathbf{\gamma }$\ vector is built in terms of the $\alpha
_{l}$'s\ as,
\begin{equation}
\mathbf{\gamma =}\left( \alpha _{l}.\alpha _{0}\right) ^{-1}\alpha
_{l};\qquad \mathbf{\gamma .}\alpha _{0}=1
\end{equation}
This relation is dictated from eqs(6.55-56), which shows the the Larange
field parameter $\Lambda $ can also be written as
\begin{equation}
\Lambda =-\mathbf{\gamma .b}_{0}
\end{equation}
Now using eqs(6.60-62), the effective field action reads, at an any order of
$\theta $, as follows
\begin{eqnarray}
S\left[ \mathbf{b}_{\mu },\mathcal{A}_{\mu }^{ex}\right] &=&\frac{\mathrm{1}%
}{4\pi }\int d^{3}y\varepsilon ^{\mu \nu \sigma }\left[ \mathbf{b}_{\mu
}\ast \left( \alpha _{0}\mathbf{\partial }_{\nu }+\frac{2i}{3}\mathbf{\gamma
.b}_{\nu }\right) \ast \mathbf{b}_{\sigma }\right]  \notag \\
&&-\frac{\mathrm{1}}{2\pi }\int d^{3}y\varepsilon ^{\mu \nu \sigma }\left[
\mathbf{q}.\mathbf{b}_{\mu }\ast \alpha _{0}\mathbf{\partial }_{\nu }%
\mathcal{A}_{\sigma }^{ex}\right]
\end{eqnarray}
where $\ast $ is the usual Moyal product.

\section{Discussion and Conclusion}

\qquad So far we have studied the NC effective field and matrix models for
FQH states with rational filling factors. These states come either from: (1)
a multilayer system $\left\{ \cup _{a}\mathcal{L}_{a}\right\} $ of Laughlin
type; that is the $\mathcal{L}_{a}$'s have individual filling factors $%
1/k_{a}$ or (2) as a singular layer hierarchical states. The latter were
interpreted as describing branches of the Hall fluid and were realized in
terms of fiber bundles on the layer $\mathcal{L}$. What remains to do is to
complete this study by giving the case where FQH states with rational
filling factors that come from multilayer hierarchical systems; see figure
11. This is not a problem since we have all the necessary tools to study
this question.

\begin{figure}[h]
\vspace{1.5cm}
\begin{equation*}
\epsfxsize=9cm \epsffile{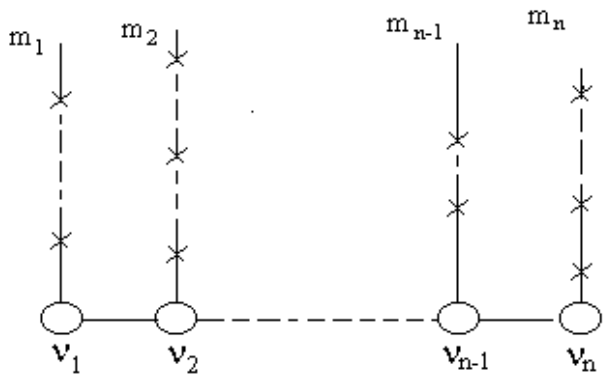}
\end{equation*}
\caption{\textit{This figure represents a specific FQH system of }$n$\textit{%
\ parallel layers }$\{\mathcal{L}_{a}\}$\textit{\ with closest interactions.
Each layer} $\mathcal{L}_{a}$\textit{\ has} $m_{a}$ {occupied Landau levels.
The filling factor is usually given by a formula type eq(3.53)}\textit{. For
layers with different individual filling factors }$\protect\nu _{a}$;
\textit{the symmetry of the model is} $U\left( 1\right) ^{n}$; \textit{while
for the case where some of the} $\protect\nu _{a}$'\textit{s are equal, one
may have non abelian subgroups of} $U\left( n\right) $. \textit{More general
FQH states involve obviously more complicated simplex.}}
\label{figure 11}
\end{figure}

To that purpose, we start from eq(6.63) describing the effective field model
for a single layer hierarchical states and implement the non abelian
structure induced from the base {\large B} consisting of $n$ $D2$ branes. In
the special case where $\left| d_{b}-d_{a}\right| $ goes to zero and
moreover all individual hierarchical states of the layers are identical,
\begin{equation}
\mathrm{\nu }_{m_{1}}=\mathrm{\nu }_{m_{2}}=...=\mathrm{\nu }_{m_{n}}=%
\mathrm{\nu }_{m}
\end{equation}
one has an exact non abelian $U\left( n\right) $ symmetry and so the NC
effective field gauge action follows by substituting the single layer vector
bundle $\mathbf{b}_{\mu }$ field by the following $n\times n$ matrix,
\begin{equation}
\mathbf{B}_{\mu }=\sum_{a,b=1}^{n}\mathbf{u}_{ab}\text{ }\mathbf{B}_{\mu
}^{ab}
\end{equation}
where $\mathbf{u}_{ab}$ is as in eq(2.20) and where $\mathbf{B}_{\mu }^{ab}$
are gauge fields connecting layers. These fields are also vectors of the
fiber bundle and have the decomposition
\begin{equation}
\mathbf{B}_{\mu }=\sum_{i=1}^{nm}\alpha _{i}\text{ }\mathbf{B}_{\mu }^{i}
\end{equation}
where $\alpha _{i}$ is the vector basis of the fiber bundle \textsl{F}$%
\left( B\right) $ based on $\left\{ \cup _{a}\mathcal{L}_{a}\right\} $.
Taking into account both the decompositions (7.2-3), we can also expand $%
\mathbf{B}_{\mu }$ as
\begin{equation}
\mathbf{B}_{\mu }=\sum_{a,b=1}^{n}\sum_{i=1}^{nm}\alpha _{i}\text{ }b_{\mu
}^{iab}\text{\ }\mathbf{u}_{ab}
\end{equation}
where $b_{\mu }^{iab}$\ are the gauge components on the vector bundle
\textsl{F}$\left( \cup _{a=1}^{n}\mathcal{L}_{a}\right) $, see eq(2.19). The
generalized intersection matrix extending the single layer matrix of
eq(6.1), which correspond to setting $n=m=1$, reads as
\begin{equation}
\alpha _{i}.\alpha _{j}=\widetilde{\mathrm{G}}_{ij};\quad i,j=1,...,nm
\end{equation}
With these fields and intersection matrix, one can write down the
generalized NC effective gauge field model for hierarchical states living on
the multilayer system. In the case all layers have the same filling factor,
the action of the model has a non abelian $U\left( n\right) $ gauge
invariance and reads in terms of the $\mathbf{B}_{\mu }^{i}$ component
fields as
\begin{eqnarray}
S\left[ \mathbf{B}_{\sigma }^{m},\mathcal{A}_{\mu }^{ex}\right] &=&\frac{%
\mathrm{1}}{4\pi }\int d^{3}y\varepsilon ^{\mu \nu \sigma }\text{ }%
\widetilde{\mathrm{G}}_{ij}\mathrm{Tr}_{U\left( n\right) }\left[ \mathbf{B}%
_{\mu }^{i}\ast \partial _{\nu }\mathbf{B}_{\sigma }^{j}+\frac{2i}{3}%
\widetilde{\mathrm{G}}_{0l}^{-1}\widetilde{\mathrm{G}}_{lk}\text{ }\mathbf{B}%
_{\mu }^{i}\ast \mathbf{B}_{\nu }^{k}\ast \mathbf{B}_{\sigma }^{j}\right]
\notag \\
&&-\frac{\mathrm{1}}{2\pi }\int d^{3}y\varepsilon ^{\mu \nu \sigma }\mathrm{%
Tr}_{U\left( n\right) }\left[ \mathbf{q}_{i}.\mathbf{B}_{\mu }^{i}\ast
\alpha _{0}\mathbf{\partial }_{\nu }\mathcal{A}_{\sigma }^{ex}\right]
\end{eqnarray}
Like for eq(6.63), this action may be rewritten in a more condensed form,
without using the expansion (7.3), as
\begin{eqnarray}
S\left[ \mathbf{B}_{\mu },\mathcal{A}_{\mu }^{ex}\right] &=&\frac{\mathrm{1}%
}{4\pi }\int d^{3}y\varepsilon ^{\mu \nu \sigma }\mathrm{Tr}_{U\left(
n\right) }\left[ \mathbf{B}_{\mu }\ast \left( \alpha _{0}\mathbf{\partial }%
_{\nu }+\frac{2i}{3}\mathbf{\gamma .B}_{\nu }\right) \ast \mathbf{B}_{\sigma
}\right]  \notag \\
&&-\frac{\mathrm{1}}{2\pi }\int d^{3}y\varepsilon ^{\mu \nu \sigma }\mathrm{%
Tr}_{U\left( n\right) }\left[ \mathbf{q}.\mathbf{B}_{\mu }\ast \alpha _{0}%
\mathbf{\partial }_{\nu }\mathcal{A}_{\sigma }^{ex}\right] ,
\end{eqnarray}
where $\mathbf{\partial }_{\nu }$\ and $\mathbf{\gamma }$ are as in
eqs(6.60-62). Since in the above $U\left( n\right) $ symmetric model, all
layer have same individual filling factor $\mathrm{\nu }_{m}$, the total
filling factor of this system is,
\begin{equation}
\mathrm{\nu }_{tat}=n\mathrm{\nu }_{m}.
\end{equation}
It extends naturally the case $\mathrm{\nu }_{tat}=n\mathrm{\nu }_{1}=n/k$
eq(1.4), where all the layers of the system are in the first Landau level;
that is of Laughlin type. Along with this $U\left( n\right) $\ maximal
symmetric situation, one can also derive\ the other phases of the multilayer
system corresponding to the situations where the maximal $U\left( n\right) $
gauge group is broken down to subgroups. Though a little bit cumbersome,
these phases are easily computed by following the same lines of the analysis
that we have given in sections 3 and 4.

To conclude, we have studied in this paper NC effective field and matrix
models for FQH states with rational filling factors. After having given a
classification of such states, we have presented our way in handling
multilayer and hierarchical states with rational filling factors by using
the correspondences: one layer $\mathcal{L}_{a}\longleftrightarrow $ one $D2$
brane and thinking about Landau levels $\left\{ \QTR{sl}{LL}_{i}\right\} $
as fiber bundles on these $D2$ branes; i.e
\begin{eqnarray}
\mathcal{L}_{a} &\longleftrightarrow &D2\longleftrightarrow {\large B}=\text{%
Base manifold,}  \notag \\
\QTR{sl}{LL}_{i} &\longleftrightarrow &\QTR{sl}{F}\left( D2\right)
\longleftrightarrow \QTR{sl}{V}=\text{ Fiber bundle}
\end{eqnarray}
Then, we have studied several representations of FQH states with rational
filling factors. These realizations are mainly given by the three following:

(a) Multilayer FQH states where all layers are taken in the first Landau
levels. Here we have shown that solutions preserving gauge invariance
constraint eqs are specified by the $\left| d_{b}-d_{a}\right| $ distances
separating branes; but also by the individual filling factors $\mathrm{\nu }%
_{a}$. Although the distances $\left| d_{b}-d_{a}\right| $ go to zero $%
\forall a,b$, $U\left( n\right) $ invariance is not automatic; it requires
moreover that all individual filling factors $\mathrm{\nu }_{a}$ have to be
equal. This remarkable property has allowed us to distinguish between
various kinds of effective fields models; in particular those having abelian
gauge symmetries with and without interactions. In section 3, we have
studied examples of these solutions and derived new kinds of series, such
that
\begin{equation}
\mathrm{\nu }=\frac{k_{1}+k_{2}}{k_{1}k_{2}-l^{2}};\quad k_{1},k_{2}\text{
odd integers,}
\end{equation}
containing as special cases the level two Jain and Haldane sequences.

This study has also permitted us to shed more light on the varieties of
generalizations of Haldane wave functions one encounters in FQH literature.
We have taken the opportunity to propose a classification of the
generalizations of Laughlin wave function. To our knowledge neither the wave
function generalizing the Laughlin wave function with maximal $U\left(
n\right) $ gauge symmetry we have obtained in the present study eq(3.61),
nor the classification of their generalizations had been known before.
Commun results in this matter are naturally covered in our present analysis.

(b) Motivated by the natural correspondence (7.9), we have revisited Wen-Zee
effective field model used to describe hierarchical states. First we have
shown explicitly that the level of the hierarchy is also the number of the
lowest Landau levels $\left\{ \QTR{sl}{LL}_{i}\right\} $ and, as far as
Haldane and Jain series are considered, each level of the hierarchy may be
though of as described by a kind of Laughlin type state. Though this
property is quite manifest on the Jain models; it is however no so obvious
for Haldane states. In this regards, we have shown, using an exact
mathematical result, that Haldane series may be usually decomposed as a sum
over Laughlin type states. To our knowledge such a feature was checked only
through specific examples such as $2/5=1/3+1/15$ and $3/7=2/5+1/35$. In
subsection 5.3, we have given the exact result to any order; see equations
(5.21-24). After having built explicit realizations of Jain and Haldane
fiber bundles and shown, amongst others, that Wen-Zee gauge field model
follows in straightforward way using fibration ideas, we have studied NC
effective field and matrix models extending the Wen-Zee commutative one and
recovering the Susskind NC field and Susskind Polychronakos matrix models as
special cases.

(c) Combining single layer hierarchy and multilayer states representations,
one gets the general situation of FQH fluids. We have also studied this
representation and have shown that there are several cases depending on the
layers spacings, gauge symmetry and individual filling factors. The most $%
U\left( n\right) $ symmetric solution one can imagine, has been
discussed in the beginning of this section. The general result for
subgroups of $U\left( n\right) $ follows naturally from the
discussion we have given in this section and the study made in
sections 3, 4 and 6.
\section*{ Acknowledgments}
 EL Hassan Saidi would like
to thank Ahmed Jellal, Hendrik Geyer and El Mostapha Sahraoui for
earlier collaborations on this FQH issues.

\end{document}